\documentclass[%
 reprint,
 amsmath,amssymb,
 aps,
prl,
]{revtex4-2}

\usepackage{graphicx}
\usepackage{dcolumn}
\usepackage{bm}

\usepackage[utf8]{inputenc}
\usepackage[T1]{fontenc}
\usepackage{mathptmx}
\usepackage{etoolbox}
\usepackage{xr}

\makeatletter
\def\@email#1#2{%
 \endgroup
 \patchcmd{\titleblock@produce}
  {\frontmatter@RRAPformat}
  {\frontmatter@RRAPformat{\produce@RRAP{*#1\href{mailto:#2}{#2}}}\frontmatter@RRAPformat}
  {}{}
}%
\makeatother

\graphicspath{
	{figures/}
}

\begin{document}

\title{Attosecond Diffraction Imaging of Electron Dynamics in Solids}

\author{Mingrui Yuan}
\affiliation{Wyant College of Optical Sciences, University of Arizona, Tucson, AZ 85721}
\affiliation{Department of Physics, University of Arizona, Tucson, AZ 85721}

\author{Nikolay V. Golubev}
\email{ngolubev@arizona.edu}
\affiliation{Department of Physics, University of Arizona, Tucson, AZ 85721}

\date{\today}

\begin{abstract}
Visualizing the electron dynamics in four dimensions of space and time is crucial to the understanding of several ubiquitous processes in nature. Hence, ultrafast X-ray and electron imaging tools have been developed to probe the dynamics of matter by means of the time-resolved diffraction imaging (TRDI). In this work, we report an extension of the theory underlying the TRDI to the case of the laser-driven electron dynamics in solid state systems. We demonstrate that the TRDI signal encodes essential information about the time-dependent electron density of the system under study and thus makes it possible to decipher the ultrafast quantum dynamics and the electron transfer phenomena in solids. We apply the developed approach to image the laser-driven electron dynamics in neutral graphene showing that the predictions made by the fully quantum version of the TRDI deviate significantly from those obtained with the conventional semiclassical approach.
\end{abstract}

\pacs{Valid PACS appear here}
\maketitle

Recent advancements in attosecond technologies~\cite{krausz2009} have facilitated the observation and eventually control of ultrafast electron dynamics in atoms~\cite{goulielmakis2010}, clusters~\cite{gong2022}, molecules~\cite{calegari2014,kraus2015,matselyukh2022}, liquids~\cite{jordan2020}, and solids~\cite{borrego-varillas2022,hui2022} with unprecedented resolution. However, the majority of previously utilized techniques, such as high-harmonic generation~\cite{kraus2015}, attosecond transient absorption~\cite{matselyukh2022}, photoelectron~\cite{gruson2016} and photofragmentation~\cite{sansone2010,calegari2014} spectroscopies, are specifically designed to capture the temporal evolution of a system but often provide insufficient information about the electron dynamic in the space domain.

Established experimental techniques that have been used for over a century to image the spatial structure of matter are X-ray~\cite{cowley1995,schlichting2012} and electron diffraction~\cite{colliex2006,carter2016,Hassan2017,Hassan2018} spectroscopies. With the advent of free-electron lasers~\cite{ishikawa2012,pellegrini2016} it is now possible to generate short coherent pulses in X-ray energy domain permitting thus direct observations of dynamics of matter with atomic spatial resolution. Even better temporal resolution has been recently demonstrated in attosecond electron diffraction microscopy experiment measuring the electron dynamics in the neutral multilayer graphene~\cite{hui2023}. This attomicroscopy setup together with the free-electron laser facilities provide access to long-awaited real-space imaging of electron motion in matter on its natural time scales.

Theoretical description of X-ray and electron scattering from matter has been widely studied before~\cite{coppens1992,coppens1997}. A traditional semiclassical (SC) approach to connect the scattering signal with the properties of a system under study is based on a simple relation between the electron density $Q(\mathbf{r})$, where $\mathbf{r}$ denotes the real-space coordinate, and the intensity of the scattered beam $I(\mathbf{S})$, where $\mathbf{S}$ is the scattering vector. Under the first Born approximation and assuming the scattering is elastic, the scattering intensity is proportional to the absolute value of the Fourier transform of the electron density~\cite{coppens1992,coppens1997}: $I(\mathbf{S}) \propto |\hat{\mathcal{F}}_{\mathbf{S}}[Q(\mathbf{r})]|^2$, where $\hat{\mathcal{F}}$ denotes the Fourier transform operator. While in the past the computation of the electron density of a system $Q(\mathbf{r})$ was a complex problem by itself, the ever growing computational power made the simulations of the ground state electron density a routine task which, in turn, permitted the high-accuracy calculations of the scattering signal from complex systems.

Extending the SC approach to the case of a time-dependent target, one could assume that the intensities of the scattered beams will reflect the dynamics of the electron density in a similar manner~\cite{cao1998a,authier2008,centurion2022}: $I(\mathbf{S},t) \propto |\hat{\mathcal{F}}_{\mathbf{S}}[Q(\mathbf{r},t)]|^2$. Indeed, if the duration of the probe pulse is much shorter than the time scale of the electron motion underlying the evolution of the electron density $Q(\mathbf{r},t)$, the SC formalism seems to be a natural choice. The SC approach has been widely used in the recent past to describe the time-resolved diffraction imaging (TRDI) in atoms~\cite{shao2010,suominen2014}, molecules~\cite{shao2010,baum2010}, and solids~\cite{yakovlev2015}.

A more careful consideration of the physics behind the time-resolved scattering process revealed fundamental issues with the applicability of the SC approach to the case of the scattering from a time-dependent system. In their pioneering work~\cite{dixit2012}, Dixit and co-workers utilized a consistent description of light-matter interaction in the frame of quantum electrodynamics to demonstrate that the time-resolved scattering cross section deviates significantly from its SC analogue. It has been shown~\cite{dixit2012,dixit2014} that the scattering signal at particular time instance $t$ is related to a special form of the so-called density-density correlation function $\langle \Psi(t+\tau/2)| \hat{\rho}(\mathbf{r}') e^{-i\hat{H}\tau} \hat{\rho}(\mathbf{r})| \Psi(t-\tau/2) \rangle$, where $|\Psi\rangle$ denotes the time-dependent wavefunction of the system evolving according to the Hamiltonian $\hat{H}$, $\hat{\rho}(\mathbf{r})$ and $\hat{\rho}(\mathbf{r}')$ are density operators, and the parameter $\tau$ accounts for the propagation of the probe pulse through the system. This fully quantum mechanical (QM) approach to the TRDI, QM-TRDI, has been applied to probe the pure electronic dynamics in atomic~\cite{dixit2012,dixit2013,dixit2013a,shao2013,dixit2014,shao2014,shao2016,dixit2017,grosser2017,shao2017} and molecular systems~\cite{bennett2014,hermann2020,giri2021,rouxel2021,tremblay2021}, and also non-adiabatic molecular transitions at avoided crossings and conical intersections~\cite{kowalewski2017,bennett2018,simmermacher2019,giri2022,tremblay2023a}. Interestingly, the applications of the QM-TRDI to solid state systems have not been reported yet. This is especially surprising in view of the fact that the X-ray diffraction imaging has been initially invented~\cite{vonLaue1912} and then very successfully used~\cite{jones2014} to investigate the structure and properties of crystals for more than a hundred years.

In this Letter, we extend the theory underlying the QM-TRDI to the case of the laser-driven electron dynamics in solid state systems. We demonstrate that the simulations based on the QM-TRDI deviate substantially from that performed with more conventional SC-TRDI approach. We apply the developed formalism to probe the ultrafast laser-driven electron dynamics in neutral graphene, showing the possibility to trace the dynamics of the electron density, and thus demonstrate that the QM-TRDI can be a very useful technique to study ultrafast electron motion in solids.

The choice of graphene is motivated by several reasons. First of all, graphene is one of the simplest solids which makes it computationally affordable to perform fully quantum modeling of the laser-driven electron dynamics and the subsequent simulations of the diffraction signal. Second, graphene has been extensively studied theoretically and thus both the electronic structure~\cite{castroneto2009} and the laser dynamics~\cite{ishikawa2010,kelardeh2015,liu2018,morimoto2022} in this system are well understood. Furthermore, Yakovlev and co-workers demonstrated the application of the SC-TRDI to the case of graphene and identified certain characteristic features in their calculated diffraction signal~\cite{yakovlev2015}. Finally, and this is our third reason to study graphene, the electron dynamics in this system has been recently recorded experimentally by means of the attosecond electron diffraction~\cite{hui2023}. Importantly, the measured diffraction signal contradicts with the SC-TRDI simulations made by Yakovlev \textit{et al.}~\cite{yakovlev2015}. Therefore, graphene turns out to be a perfect system to test the validity of the QM-TRDI approach and explain the difference between the previous theoretical simulations and the experimentally measured results.

Graphene is a monolayer material composed of carbon atoms arranged in a two-dimensional hexagonal honeycomb lattice structure~\cite{castroneto2009}. Each carbon atom of graphene possesses six electrons while four of them participate in the formation of valence bonds. Three of these (the so-called $\sigma$-electrons) form tight bonds with the neighboring atoms in the plane while the fourth electron participates in the formation of $\pi$ (bonding) and $\pi^*$ (anti-bonding) bonds that lie above and below of the plane formed by atomic nuclei. The unique structure and interactions of these delocalized $\pi$ and $\pi^*$ orbitals of graphene made this material a holy grail of solid state physics attracting ever growing interest from the community.

The electronic structure of graphene can be described using a rather simple tight-binding model, leading to analytical solutions for the band energies, related eigenstates, and the corresponding electric dipole matrix elements connecting these states with each other. Taking only $\pi$ electrons into account, we form the two-band system consisting of the valence and the conduction bands of graphene (see Secs. I and II of the Supplemental Material (SM) for details). These bands are highly symmetric in reciprocal space and touch each other at the so-called Dirac points forming thus conical intersections~\cite{castroneto2009}. In case of graphene, the intersection points lie exactly at the Fermi level making this material a semimetal (zero gap) character and leading to an enhanced transition probability from one band to the other at these points. Accordingly, in the following we will focus on the laser-driven electron dynamics occurring in the vicinity of the Dirac points of graphene.

We calculated the evolution of the electron density in reciprocal space of graphene driven by the interaction of the system with a laser field of the following waveform: $\mathbf{E}(t)=\mathbf{e} E_0 \sin^4(\pi t/\tau)\cos(\omega t)$, where the peak field amplitude $E_0$ is chosen to be 2.5~V/nm, pulse duration $\tau$ is 21~fs, the photon energy $\omega$ is 1.55~eV which corresponds to the wavelength of 800~nm, and $\mathbf{e}$ denotes the unit vector in the direction of the field polarization. The chosen laser pulse parameters are comparable to those utilized in the recent experiment Ref.~\cite{hui2023}. We numerically solved the semiconductor Bloch equation employing the dipole approximation and utilizing the time-dependent crystal momentum frame evolving according to the Bloch acceleration theorem~\cite{houston1940,krieger1986}: $\mathbf{k}_t=\mathbf{k}_0 + e/\hbar \mathbf{A}(t)$, where $e$ is the electron charge, $\mathbf{k}_t$ and $\mathbf{k}_0$ denote the time-dependent and the field-free reciprocal space vectors, respectively, and $\mathbf{A}(t)=-\int_{-\infty}^t \mathbf{E}(t') dt'$ is the vector potential associated with the applied laser pulse $\mathbf{E}(t)$. The detailed description of the employed computational procedure can be found in Sec.~V of SM.

\begin{center}
\begin{figure*}[t]
	\includegraphics[width=0.98\textwidth]{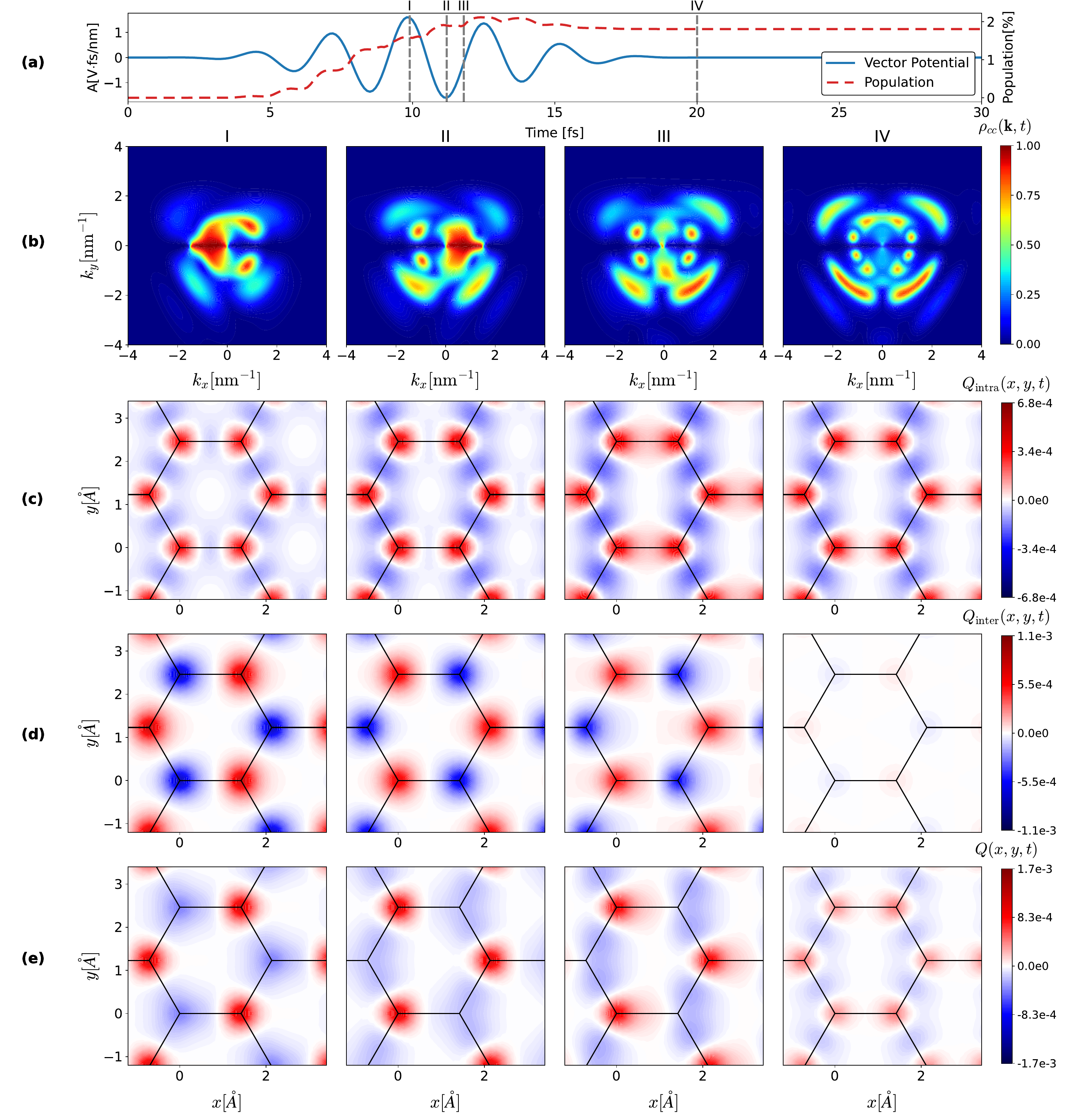}
\caption{Light-driven electron dynamics in graphene. (a) The blue curve represents the vector potential of the pump laser field, while the red line depicts the evolution of the population of the conduction band of graphene. (b) The snapshots of the electron density distribution in the reciprocal space of graphene in the vicinity of the Dirac point placed at the origin. The four columns depict the time instances $t_I=9.8$~fs, $t_{II}=11.2$~fs, $t_{III}=11.8$~fs, and $t_{IV}=20.0$~fs, indicated by the vertical grey dashed lines in (a). (c), (d) and (e) The corresponding integrated real space intraband, interband, and the total electron difference densities, respectively, for the same time instances shown in (a) and (b).}
\label{fig:graphene_dynamics}
\end{figure*}
\end{center}

The action of the external field on graphene causes population transfer between the valence and conduction bands via two distinguishable mechanisms: (i) direct photon absorption occurring in the regions of reciprocal space where the energy gap between the bands matches the photon energy $\omega$, and (ii) the so-called Landau--Zener transitions~\cite{higuchi2017,ivakhnenko2023} resulting from the transfer of electrons through conical intersections in the vicinity of Dirac points. Figure~\ref{fig:graphene_dynamics} summarizes the results of the electron dynamics driven by the laser pulse described above and which is polarized along C--C bonds of the graphene sample ($x$ direction in our simulations). The panel (a) of Fig.~\ref{fig:graphene_dynamics} depicts the vector potential $\mathbf{A}(t)$ (blue solid line) and the corresponding population of the conduction band of graphene integrated over the unit cell in the reciprocal space (red dashed line). As one can see, the population dynamics occurs continuously along the action of the field leading to about 2\% electron transfer between the valence and conduction bands.

The panel (b) of Fig.~\ref{fig:graphene_dynamics} shows the snapshots of the matrix element $\rho_{c,c}(\mathbf{k},t)$, where the index $c$ denotes the conduction band, of the reciprocal space density matrix $\bm{\rho}(\mathbf{k},t)$ in the vicinity of the Dirac point located at the origin. It is seen that the electron density is displaced towards negative values of $k_x$ direction when the vector potential is positive (snapshot I in Fig.~\ref{fig:graphene_dynamics}(b), taken at $t_{I}=9.8$~fs as indicated by the vertical dashed line in the panel (a)) and towards positive values of $k_x$ when the sign of the vector potential is reversed (snapshot II, $t_{II}=11.2$~fs). Importantly, at the moments of time when the vector potential is zero either during the action of the field (snapshot III, $t_{III}=11.8$~fs) or when the field is over (snapshot IV, $t_{IV}=20$~fs), the electron density in the reciprocal space remains symmetric with respect to the vertical line passing through the conical intersection.

The calculated reciprocal space density matrix $\bm{\rho}(\mathbf{k},t)$ gives access to the real space time-dependent electron density $Q(\mathbf{r},t)$:
\begin{equation}
\label{eq:Q}
	Q(\mathbf{r},t)=\frac{1}{\mathcal{N}} \sum_{n,m} \int_{\text{unit cell}} \rho_{n,m}(\mathbf{k},t) Q_{m,n} (\mathbf{k}_t,\mathbf{r}) d \mathbf{k},
\end{equation}
where the indices $n$ and $m$ iterate over the valence ($v$) and conduction ($c$) bands, the integration is performed over the unit cell in the reciprocal space, $\mathcal{N}$ is the normalization factor accounting for the volume of the unit cell and the number of active electrons ($\mathcal{N}=(2\pi)^2/\sqrt{3}a^2$ for graphene, where $a=2.46 \AA$ is the lattice constant), and $Q_{m,n} (\mathbf{k},\mathbf{r})$ denote the $\mathbf{k}$- and $\mathbf{r}$-resolved electron densities of $\pi$/$\pi^*$ bands ($m=n=v/c$, respectively), or the corresponding transition densities connecting the bands with each other ($m \neq n$). The total real-space electron density $Q(\mathbf{r},t)$ can be decomposed in two contributions: $Q(\mathbf{r},t)=Q_{\text{intra}}(\mathbf{r},t)+Q_{\text{inter}}(\mathbf{r},t)$, where the first term $Q_{\text{intra}}(\mathbf{r},t)$ represents the diagonal portion of Eq.~(\ref{eq:Q}) ($n=m\in\{v,c\}$) and thus is referred to as the intraband electron density, while the second term $Q_{\text{inter}}(\mathbf{r},t)$ is composed from the off-diagonal elements responsible for the coherent interband electron motion (see Sec.~I of SM for the detailed derivation). In the follow-up discussion we will use the real space electron difference densities (EDDs) integrated over $z$ axis for convenience: $Q(x,y,t)=\int [Q(\mathbf{r},t) - Q(\mathbf{r},0)] dz$, where we removed, in addition, the static portion of the density to make the corresponding figures more representative.

The panels (c), (d), and (e) of Fig.~\ref{fig:graphene_dynamics} depict the snapshots of the intraband $Q_{\text{intra}}(x,y,t)$, interband $Q_{\text{inter}}(x,y,t)$, and the total $Q(x,y,t)$ EDDs for the same time instances I, II, III, and IV discussed before. As one can see, the intraband EDDs for time instances I and II (columns I and II in Fig.~\ref{fig:graphene_dynamics}(c)) are nearly identical to each other up to a scaling factor reflecting the difference in the overall population of the conduction band. At the same time, the corresponding reciprocal space densities (Fig.~\ref{fig:graphene_dynamics}(b), I and II) mirror each other with respect to the $y$ axis passing through the origin as we already discussed in the previous paragraph. The identical real-space EDDs for the distinguishable reciprocal space densities arise due to the symmetry of the $Q_{m,n} (\mathbf{k},\mathbf{r})$ densities with respect to the Dirac point in graphene. In a difference, the corresponding interband EDDs, shown in Fig.~\ref{fig:graphene_dynamics}(d), I and II, are antisymmetric with respect to each other. Accordingly, the total EDDs for time instances I and II (Fig.~\ref{fig:graphene_dynamics}(e), I and II), which are combinations of the intra- and interband contributions, are also antisymmetric yet qualitatively identical to each other. The intraband EDDs for time instances III and IV are also similar to each other (Fig.~\ref{fig:graphene_dynamics}(c), III and IV), same as the corresponding reciprocal space densities (Fig.~\ref{fig:graphene_dynamics}(b), III and IV). We note, however, a significant difference between the intraband EDDs and the reciprocal space densities for time instances I/II and III/IV, respectively. The interband EDD at time instance IV (Fig.~\ref{fig:graphene_dynamics}(d), IV) has nearly disappeared due to the fast decoherence of the matrix element $\rho_{vc}(\mathbf{k},t)$ and its complex conjugate. As the result, the total EDD is composed almost entirely from the intraband component shortly after the action of the pulse on the system is over (see Fig.~\ref{fig:graphene_dynamics}(e), IV). The complete movie of the laser-driven electron dynamics in graphene can be found in SM Video I.

\begin{figure}[t]
	\includegraphics[width=8.5cm]{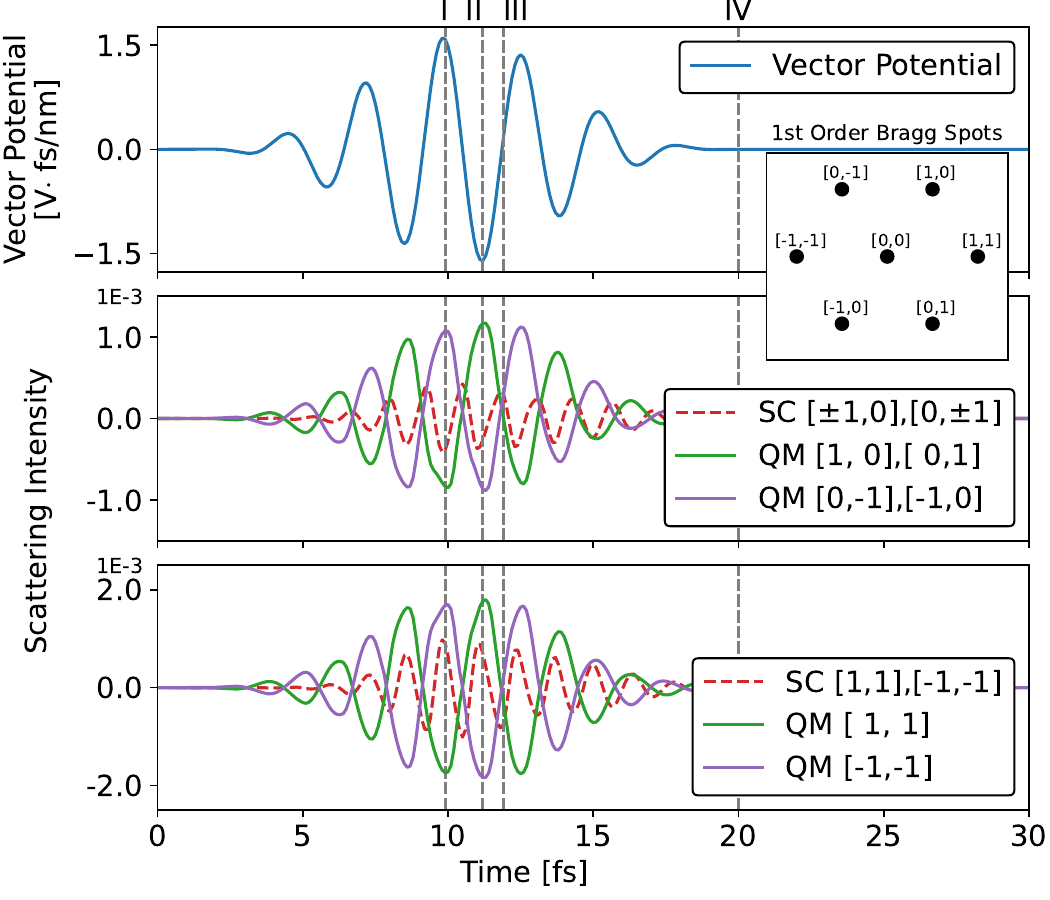}
\caption{Comparison of the time-resolved Bragg diffraction intensities in graphene computed with the semiclassical (SC, red dashed lines) and quantum mechanical (QM, green and purple solid lines) approaches. Top panel: Vector potential of the pump field (blue solid line). Middle and bottom panels: SC and QM diffraction intensities measured at $[\pm1,0]$, $[0,\pm1]$ and $[1,1]$, $[-1,-1]$ Bragg spots, respectively. Note that SC intensity is scaled by a factor of four for better visibility. The vertical grey dashed lines indicate the four time instances depicted in Fig.~\ref{fig:graphene_dynamics}. The inset schematically shows positions of the zero- and first-order Bragg spots of graphene.}
\label{fig:diffraction}
\end{figure}

Utilizing the formalism discussed above, the intensities of Bragg diffraction peaks in the SC-TRDI approximation can be expressed as
\begin{equation}
\label{eq:SC_TRDI}
\begin{split}
	I_{\text{SC}} & (\mathbf{S},t) = \left| \mathcal{\hat{F}}_{\mathbf{S}}[Q(\mathbf{r},t)] \right|^2 \\
	& =\frac{1}{\mathcal{N}} \left|
		\sum_{n,m}
		\int_{\text{unit cell}} \rho_{n,m}(\mathbf{k},t)
			\mathcal{\hat{F}}_{\mathbf{S}}[Q_{m,n} (\mathbf{k}_t,\mathbf{r})]
		d\mathbf{k}
		\right|^2,
\end{split}
\end{equation}
where the scattering vector $\mathbf{S}=h \frac{2\pi}{a}(1/\sqrt{3},1)+l \frac{2\pi}{a}(1/\sqrt{3},-1)$ is composed from the corresponding reciprocal space vectors of graphene, and $h$ and $l$ are integers (see Sec.~III of SM for details). We simulated the SC-TRDI for the first-order Bragg peaks of graphene. Due to the high symmetry of the graphene structure, which is reflected in the corresponding symmetries of the Fourier transforms $\mathcal{\hat{F}}_{\mathbf{S}}[Q_{m,n} (\mathbf{k}_t,\mathbf{r})]$ contributing to Eq.~(\ref{eq:SC_TRDI}), only two signals $I_{\text{SC}}(\mathbf{S}_1,t)$ ($[h=1,l=0]$, $[0,1]$, $[0,-1]$, $[-1,0]$) and $I_{\text{SC}}(\mathbf{S}_2,t)$ ($[1,1]$, $[-1,-1]$) are distinguishable from each other. Figure~\ref{fig:diffraction} depicts the correspondence between the applied vector potential (blue solid line in the top panel) and the simulated ``normalized'' (excluding the static background and the part responsible for the population transfer) SC-TRDI signal for the two families of Bragg spots $I_{\text{SC}}(\mathbf{S}_1,t)$ and $I_{\text{SC}}(\mathbf{S}_2,t)$, shown by the red dashed lines in the middle and bottom panels, respectively. As one can see, the diffraction intensities at the maximum ($t_{I}=9.8$~fs) and minimum ($t_{II}=11.2$~fs) values of the vector potential have the same sign and comparable magnitudes with each other. This is again do to the symmetry of the Fourier transform: $\left| \mathcal{\hat{F}}_{\mathbf{S}}[Q(\mathbf{r},t)] \right|^2 = \left| \mathcal{\hat{F}}_{\mathbf{S}}[Q(-\mathbf{r},t)] \right|^2$. The total EDDs, as well as their components, are either fully symmetric ($Q_{\text{intra}}(x,y,t)$, Fig.~\ref{fig:graphene_dynamics}(c), I and II) or antisymmetric ($Q_{\text{inter}}(x,y,t)$ and $Q(x,y,t)$, Fig.~\ref{fig:graphene_dynamics}(d) and (e), I and II) with respect to each other at time instances I and II. Accordingly, the absolute values of their Fourier transforms are comparable to each other which makes the SC-TRDI signal oscillating twice the frequency of the applied external field.
 
Let us turn to the simulations employing the QM-TRDI approach. Assuming the duration of the probe pulse is short and that the photon energy transfer during the scattering event is small, such that $\omega_{\mathbf{k}_{\text{s}}} \approx \omega_{\mathbf{k}_{\text{in}}}$, the QM-TRDI intensities can be calculated as (see Sec.~IV of SM for details)
\begin{widetext}
\begin{equation}
\label{eq:QM_TRDI}
	I_{\text{QM}}(\mathbf{S},t)=
		\frac{1}{\mathcal{N}}
		\sum_{n,m,f}
		\int_{\text{unit cell}} \rho_{n,m}(\mathbf{k},t)
			\mathcal{\hat{F}}^{*}_{\mathbf{S}}[Q_{f,m} (\mathbf{k}_t,\mathbf{r})]
			\mathcal{\hat{F}}_{\mathbf{S}}[Q_{f,n} (\mathbf{k}_t,\mathbf{r})]
		d\mathbf{k},
\end{equation}
\end{widetext}
where the index $f$ runs over the final electronic states reached by the scattering process. As one can see from Eq.~(\ref{eq:QM_TRDI}), the resulting signal depends linearly on the reciprocal space electron densities $\rho_{n,m}(\mathbf{k},t)$, while in the case of the SC-TRDI, Eq.~(\ref{eq:SC_TRDI}), the intensity is obtained as an absolute value of the sum of the corresponding quantities. Therefore, the negative (Fig.~\ref{fig:graphene_dynamics}(b), I) and positive (Fig.~\ref{fig:graphene_dynamics}(b), II) displacements of the electron density in the reciprocal space become spectroscopically distinguishable from each other in the frame of the QM-TRDI approach. Furthermore, the linearity of Eq.~(\ref{eq:QM_TRDI}) makes the total diffraction signal a simple sum of the intra- and interband contributions which is not the case for the SC-TRDI, Eq.~(\ref{eq:SC_TRDI}) (see Sec.~V of SM for the detailed discussion). The calculated normalized QM-TRDI intensities for all first-order Bragg peaks of graphene are shown in the middle and bottom panels of Fig.~\ref{fig:diffraction}. In contrast to the SC-TRDI results, the QM-TRDI signals oscillate with the same frequency as the applied external field. Furthermore, the intensities measured at spots located opposite to each other in space oscillate out of phase with respect to each other.

The major physical effect not captured by the SC-TRDI approach is, ironically, the motion of electrons, which is only partially reflected in the time-dependence of the corresponding electron density. In the SC treatment of diffraction, the beam measures an instantaneous shape of the electron density in the real space. However, two identical real-space electron densities can have different underlying electron momenta making them distinguishable in the scattering process. The QM-TRDI approach is sensitive not only to the instantaneous positions of the electrons of the target in space but also to their momenta. In our simulations, the electric field polarized along $x$ axis accelerates the electrons towards negative/positive direction when the amplitude of the corresponding vector potential is positive/negative (see panel (b) of Fig.~\ref{fig:graphene_dynamics}). Accordingly, the electrons possessing extra momentum towards certain direction increase the intensities of Bragg spots located along that direction and decrease the intensities of the opposite spots, as shown in Fig.~\ref{fig:diffraction}.

Before concluding, we emphasize that Eqs.~(\ref{eq:SC_TRDI}) and~(\ref{eq:QM_TRDI}) become identical in case of the elastic ($f=v$) scattering from the stationary target present in the ground electronic state ($\rho_{n,m}(\mathbf{k},t) = \delta_{n=m=v}$). In general, however, the QM-TRDI formalism is not equivalent to a straightforward extension of the SC approach to the case of the time-dependent electron density. Most importantly, the calculations based on the SC-TRDI approach produce the diffraction signal that oscillates twice the frequency of the driver laser field, which was already predicted earlier~\cite{yakovlev2015} (see Sec.~VII of SM for the detailed comparison of our approach and the one reported in Ref.~\cite{yakovlev2015}). Our new results demonstrate, and this agrees well with the recent experimental measurements of the electron diffraction in graphene~\cite{hui2023}, that the diffraction signal should oscillate with the same frequency as the driver field. This discrepancy between the SC and QM results arises because the SC approach ignores the momenta of the moving electrons. We would like to note that the formalism reported in our paper is not limited to simulating the electron dynamics and diffraction in a simple two-band tight-binding model of graphene and could easily be generalized to include more electronic bands into consideration or be applied to an arbitrary periodic solid (see Sec.~VI of SM for details).

In conclusion, we have demonstrated the application of QM-TRDI to probe the ultrafast laser-driven electron dynamics in solid state systems. We presented the extension of the theory underlying QM-TRDI to the case of solids and derived a simple expression that connects the dynamics of electron density in the reciprocal space with the intensities of the scattered beams. We performed fully quantum simulations of the electron dynamics in neutral graphene sample and demonstrated that QM-TRDI signal deviates significantly from that obtained with the SC-TRDI approach. Finally, we identified more complicated symmetry relations between the diffraction signals measured at different Bragg spots in comparison to those present in the SC-TRDI simulations. This, in turn, could be beneficial for the corresponding experimental measurements since more information about the dynamics of the system under study is fingerprinted in the spectra. The investigation of possible computational schemes capable of recovering the dynamics of electron density from diffraction images is a promising direction of further research. We hope that our work will motivate such studies.

\begin{acknowledgments}
The authors wish to express their sincere gratitude to Mohammed Th. Hassan for the excellent criticisms of the manuscript, and for innumerable discussions during the development of this work. The computational part of this research is based upon High Performance Computing (HPC) resources supported by the University of Arizona TRIF, UITS, and Research, Innovation, and Impact (RII) and maintained by the UArizona Research Technologies department. NVG acknowledges the financial support by the Branco Weiss Fellowship---Society in Science, administered by the ETH Z\"urich.
\end{acknowledgments}


\begin{thebibliography}{60}%
\makeatletter
\providecommand \@ifxundefined [1]{%
 \@ifx{#1\undefined}
}%
\providecommand \@ifnum [1]{%
 \ifnum #1\expandafter \@firstoftwo
 \else \expandafter \@secondoftwo
 \fi
}%
\providecommand \@ifx [1]{%
 \ifx #1\expandafter \@firstoftwo
 \else \expandafter \@secondoftwo
 \fi
}%
\providecommand \natexlab [1]{#1}%
\providecommand \enquote  [1]{``#1''}%
\providecommand \bibnamefont  [1]{#1}%
\providecommand \bibfnamefont [1]{#1}%
\providecommand \citenamefont [1]{#1}%
\providecommand \href@noop [0]{\@secondoftwo}%
\providecommand \href [0]{\begingroup \@sanitize@url \@href}%
\providecommand \@href[1]{\@@startlink{#1}\@@href}%
\providecommand \@@href[1]{\endgroup#1\@@endlink}%
\providecommand \@sanitize@url [0]{\catcode `\\12\catcode `\$12\catcode
  `\&12\catcode `\#12\catcode `\^12\catcode `\_12\catcode `\%12\relax}%
\providecommand \@@startlink[1]{}%
\providecommand \@@endlink[0]{}%
\providecommand \url  [0]{\begingroup\@sanitize@url \@url }%
\providecommand \@url [1]{\endgroup\@href {#1}{\urlprefix }}%
\providecommand \urlprefix  [0]{URL }%
\providecommand \Eprint [0]{\href }%
\providecommand \doibase [0]{https://doi.org/}%
\providecommand \selectlanguage [0]{\@gobble}%
\providecommand \bibinfo  [0]{\@secondoftwo}%
\providecommand \bibfield  [0]{\@secondoftwo}%
\providecommand \translation [1]{[#1]}%
\providecommand \BibitemOpen [0]{}%
\providecommand \bibitemStop [0]{}%
\providecommand \bibitemNoStop [0]{.\EOS\space}%
\providecommand \EOS [0]{\spacefactor3000\relax}%
\providecommand \BibitemShut  [1]{\csname bibitem#1\endcsname}%
\let\auto@bib@innerbib\@empty
\bibitem [{\citenamefont {Krausz}\ and\ \citenamefont
  {Ivanov}(2009)}]{krausz2009}%
  \BibitemOpen
  \bibfield  {author} {\bibinfo {author} {\bibfnamefont {F.}~\bibnamefont
  {Krausz}}\ and\ \bibinfo {author} {\bibfnamefont {M.}~\bibnamefont
  {Ivanov}},\ }\href {https://doi.org/10.1103/RevModPhys.81.163} {\bibfield
  {journal} {\bibinfo  {journal} {Rev. Mod. Phys.}\ }\textbf {\bibinfo {volume}
  {81}},\ \bibinfo {pages} {163} (\bibinfo {year} {2009})}\BibitemShut
  {NoStop}%
\bibitem [{\citenamefont {Goulielmakis}\ \emph {et~al.}(2010)\citenamefont
  {Goulielmakis}, \citenamefont {Loh}, \citenamefont {Wirth}, \citenamefont
  {Santra}, \citenamefont {Rohringer}, \citenamefont {Yakovlev}, \citenamefont
  {Zherebtsov}, \citenamefont {Pfeifer}, \citenamefont {Azzeer}, \citenamefont
  {Kling}, \citenamefont {Leone},\ and\ \citenamefont
  {Krausz}}]{goulielmakis2010}%
  \BibitemOpen
  \bibfield  {author} {\bibinfo {author} {\bibfnamefont {E.}~\bibnamefont
  {Goulielmakis}}, \bibinfo {author} {\bibfnamefont {Z.-H.}\ \bibnamefont
  {Loh}}, \bibinfo {author} {\bibfnamefont {A.}~\bibnamefont {Wirth}}, \bibinfo
  {author} {\bibfnamefont {R.}~\bibnamefont {Santra}}, \bibinfo {author}
  {\bibfnamefont {N.}~\bibnamefont {Rohringer}}, \bibinfo {author}
  {\bibfnamefont {V.~S.}\ \bibnamefont {Yakovlev}}, \bibinfo {author}
  {\bibfnamefont {S.}~\bibnamefont {Zherebtsov}}, \bibinfo {author}
  {\bibfnamefont {T.}~\bibnamefont {Pfeifer}}, \bibinfo {author} {\bibfnamefont
  {A.~M.}\ \bibnamefont {Azzeer}}, \bibinfo {author} {\bibfnamefont {M.~F.}\
  \bibnamefont {Kling}}, \bibinfo {author} {\bibfnamefont {S.~R.}\ \bibnamefont
  {Leone}},\ and\ \bibinfo {author} {\bibfnamefont {F.}~\bibnamefont
  {Krausz}},\ }\href {https://doi.org/10.1038/nature09212} {\bibfield
  {journal} {\bibinfo  {journal} {Nature}\ }\textbf {\bibinfo {volume} {466}},\
  \bibinfo {pages} {739} (\bibinfo {year} {2010})}\BibitemShut {NoStop}%
\bibitem [{\citenamefont {Gong}\ \emph {et~al.}(2022)\citenamefont {Gong},
  \citenamefont {Heck}, \citenamefont {Jelovina}, \citenamefont {Perry},
  \citenamefont {Zinchenko}, \citenamefont {Lucchese},\ and\ \citenamefont
  {W{\"o}rner}}]{gong2022}%
  \BibitemOpen
  \bibfield  {author} {\bibinfo {author} {\bibfnamefont {X.}~\bibnamefont
  {Gong}}, \bibinfo {author} {\bibfnamefont {S.}~\bibnamefont {Heck}}, \bibinfo
  {author} {\bibfnamefont {D.}~\bibnamefont {Jelovina}}, \bibinfo {author}
  {\bibfnamefont {C.}~\bibnamefont {Perry}}, \bibinfo {author} {\bibfnamefont
  {K.}~\bibnamefont {Zinchenko}}, \bibinfo {author} {\bibfnamefont
  {R.}~\bibnamefont {Lucchese}},\ and\ \bibinfo {author} {\bibfnamefont
  {H.~J.}\ \bibnamefont {W{\"o}rner}},\ }\href
  {https://doi.org/10.1038/s41586-022-05039-8} {\bibfield  {journal} {\bibinfo
  {journal} {Nature}\ }\textbf {\bibinfo {volume} {609}},\ \bibinfo {pages}
  {507} (\bibinfo {year} {2022})}\BibitemShut {NoStop}%
\bibitem [{\citenamefont {Calegari}\ \emph {et~al.}(2014)\citenamefont
  {Calegari}, \citenamefont {Ayuso}, \citenamefont {Trabattoni}, \citenamefont
  {Belshaw}, \citenamefont {De~Camillis}, \citenamefont {Anumula},
  \citenamefont {Frassetto}, \citenamefont {Poletto}, \citenamefont {Palacios},
  \citenamefont {Decleva}, \citenamefont {Greenwood}, \citenamefont {Martin},\
  and\ \citenamefont {Nisoli}}]{calegari2014}%
  \BibitemOpen
  \bibfield  {author} {\bibinfo {author} {\bibfnamefont {F.}~\bibnamefont
  {Calegari}}, \bibinfo {author} {\bibfnamefont {D.}~\bibnamefont {Ayuso}},
  \bibinfo {author} {\bibfnamefont {A.}~\bibnamefont {Trabattoni}}, \bibinfo
  {author} {\bibfnamefont {L.}~\bibnamefont {Belshaw}}, \bibinfo {author}
  {\bibfnamefont {S.}~\bibnamefont {De~Camillis}}, \bibinfo {author}
  {\bibfnamefont {S.}~\bibnamefont {Anumula}}, \bibinfo {author} {\bibfnamefont
  {F.}~\bibnamefont {Frassetto}}, \bibinfo {author} {\bibfnamefont
  {L.}~\bibnamefont {Poletto}}, \bibinfo {author} {\bibfnamefont
  {A.}~\bibnamefont {Palacios}}, \bibinfo {author} {\bibfnamefont
  {P.}~\bibnamefont {Decleva}}, \bibinfo {author} {\bibfnamefont {J.~B.}\
  \bibnamefont {Greenwood}}, \bibinfo {author} {\bibfnamefont {F.}~\bibnamefont
  {Martin}},\ and\ \bibinfo {author} {\bibfnamefont {M.}~\bibnamefont
  {Nisoli}},\ }\href {https://doi.org/10.1126/science.1254061} {\bibfield
  {journal} {\bibinfo  {journal} {Science}\ }\textbf {\bibinfo {volume}
  {346}},\ \bibinfo {pages} {336} (\bibinfo {year} {2014})}\BibitemShut
  {NoStop}%
\bibitem [{\citenamefont {Kraus}\ \emph {et~al.}(2015)\citenamefont {Kraus},
  \citenamefont {Mignolet}, \citenamefont {Baykusheva}, \citenamefont
  {Rupenyan}, \citenamefont {Horny}, \citenamefont {Penka}, \citenamefont
  {Grassi}, \citenamefont {Tolstikhin}, \citenamefont {Schneider},
  \citenamefont {Jensen}, \citenamefont {Madsen}, \citenamefont {Bandrauk},
  \citenamefont {Remacle},\ and\ \citenamefont {W{\"o}rner}}]{kraus2015}%
  \BibitemOpen
  \bibfield  {author} {\bibinfo {author} {\bibfnamefont {P.~M.}\ \bibnamefont
  {Kraus}}, \bibinfo {author} {\bibfnamefont {B.}~\bibnamefont {Mignolet}},
  \bibinfo {author} {\bibfnamefont {D.}~\bibnamefont {Baykusheva}}, \bibinfo
  {author} {\bibfnamefont {A.}~\bibnamefont {Rupenyan}}, \bibinfo {author}
  {\bibfnamefont {L.}~\bibnamefont {Horny}}, \bibinfo {author} {\bibfnamefont
  {E.~F.}\ \bibnamefont {Penka}}, \bibinfo {author} {\bibfnamefont
  {G.}~\bibnamefont {Grassi}}, \bibinfo {author} {\bibfnamefont {O.~I.}\
  \bibnamefont {Tolstikhin}}, \bibinfo {author} {\bibfnamefont
  {J.}~\bibnamefont {Schneider}}, \bibinfo {author} {\bibfnamefont
  {F.}~\bibnamefont {Jensen}}, \bibinfo {author} {\bibfnamefont {L.~B.}\
  \bibnamefont {Madsen}}, \bibinfo {author} {\bibfnamefont {A.~D.}\
  \bibnamefont {Bandrauk}}, \bibinfo {author} {\bibfnamefont {F.}~\bibnamefont
  {Remacle}},\ and\ \bibinfo {author} {\bibfnamefont {H.~J.}\ \bibnamefont
  {W{\"o}rner}},\ }\href {https://doi.org/10.1126/science.aab2160} {\bibfield
  {journal} {\bibinfo  {journal} {Science}\ }\textbf {\bibinfo {volume}
  {350}},\ \bibinfo {pages} {790} (\bibinfo {year} {2015})}\BibitemShut
  {NoStop}%
\bibitem [{\citenamefont {Matselyukh}\ \emph {et~al.}(2022)\citenamefont
  {Matselyukh}, \citenamefont {Despr{\'e}}, \citenamefont {Golubev},
  \citenamefont {Kuleff},\ and\ \citenamefont {W{\"o}rner}}]{matselyukh2022}%
  \BibitemOpen
  \bibfield  {author} {\bibinfo {author} {\bibfnamefont {D.~T.}\ \bibnamefont
  {Matselyukh}}, \bibinfo {author} {\bibfnamefont {V.}~\bibnamefont
  {Despr{\'e}}}, \bibinfo {author} {\bibfnamefont {N.~V.}\ \bibnamefont
  {Golubev}}, \bibinfo {author} {\bibfnamefont {A.~I.}\ \bibnamefont
  {Kuleff}},\ and\ \bibinfo {author} {\bibfnamefont {H.~J.}\ \bibnamefont
  {W{\"o}rner}},\ }\href {https://doi.org/10.1038/s41567-022-01690-0}
  {\bibfield  {journal} {\bibinfo  {journal} {Nat. Phys.}\ }\textbf {\bibinfo
  {volume} {18}},\ \bibinfo {pages} {1206} (\bibinfo {year}
  {2022})}\BibitemShut {NoStop}%
\bibitem [{\citenamefont {Jordan}\ \emph {et~al.}(2020)\citenamefont {Jordan},
  \citenamefont {Huppert}, \citenamefont {Rattenbacher}, \citenamefont {Peper},
  \citenamefont {Jelovina}, \citenamefont {Perry}, \citenamefont {Von~Conta},
  \citenamefont {Schild},\ and\ \citenamefont {W{\"o}rner}}]{jordan2020}%
  \BibitemOpen
  \bibfield  {author} {\bibinfo {author} {\bibfnamefont {I.}~\bibnamefont
  {Jordan}}, \bibinfo {author} {\bibfnamefont {M.}~\bibnamefont {Huppert}},
  \bibinfo {author} {\bibfnamefont {D.}~\bibnamefont {Rattenbacher}}, \bibinfo
  {author} {\bibfnamefont {M.}~\bibnamefont {Peper}}, \bibinfo {author}
  {\bibfnamefont {D.}~\bibnamefont {Jelovina}}, \bibinfo {author}
  {\bibfnamefont {C.}~\bibnamefont {Perry}}, \bibinfo {author} {\bibfnamefont
  {A.}~\bibnamefont {Von~Conta}}, \bibinfo {author} {\bibfnamefont
  {A.}~\bibnamefont {Schild}},\ and\ \bibinfo {author} {\bibfnamefont {H.~J.}\
  \bibnamefont {W{\"o}rner}},\ }\href {https://doi.org/10.1126/science.abb0979}
  {\bibfield  {journal} {\bibinfo  {journal} {Science}\ }\textbf {\bibinfo
  {volume} {369}},\ \bibinfo {pages} {974} (\bibinfo {year}
  {2020})}\BibitemShut {NoStop}%
\bibitem [{\citenamefont {{Borrego-Varillas}}\ \emph
  {et~al.}(2022)\citenamefont {{Borrego-Varillas}}, \citenamefont {Lucchini},\
  and\ \citenamefont {Nisoli}}]{borrego-varillas2022}%
  \BibitemOpen
  \bibfield  {author} {\bibinfo {author} {\bibfnamefont {R.}~\bibnamefont
  {{Borrego-Varillas}}}, \bibinfo {author} {\bibfnamefont {M.}~\bibnamefont
  {Lucchini}},\ and\ \bibinfo {author} {\bibfnamefont {M.}~\bibnamefont
  {Nisoli}},\ }\href {https://doi.org/10.1088/1361-6633/ac5e7f} {\bibfield
  {journal} {\bibinfo  {journal} {Rep. Prog. Phys.}\ }\textbf {\bibinfo
  {volume} {85}},\ \bibinfo {pages} {066401} (\bibinfo {year}
  {2022})}\BibitemShut {NoStop}%
\bibitem [{\citenamefont {Hui}\ \emph {et~al.}(2022)\citenamefont {Hui},
  \citenamefont {Alqattan}, \citenamefont {Yamada}, \citenamefont {Pervak},
  \citenamefont {Yabana},\ and\ \citenamefont {Hassan}}]{hui2022}%
  \BibitemOpen
  \bibfield  {author} {\bibinfo {author} {\bibfnamefont {D.}~\bibnamefont
  {Hui}}, \bibinfo {author} {\bibfnamefont {H.}~\bibnamefont {Alqattan}},
  \bibinfo {author} {\bibfnamefont {S.}~\bibnamefont {Yamada}}, \bibinfo
  {author} {\bibfnamefont {V.}~\bibnamefont {Pervak}}, \bibinfo {author}
  {\bibfnamefont {K.}~\bibnamefont {Yabana}},\ and\ \bibinfo {author}
  {\bibfnamefont {M.~T.}\ \bibnamefont {Hassan}},\ }\href
  {https://doi.org/10.1038/s41566-021-00918-4} {\bibfield  {journal} {\bibinfo
  {journal} {Nat. Photonics}\ }\textbf {\bibinfo {volume} {16}},\ \bibinfo
  {pages} {33} (\bibinfo {year} {2022})}\BibitemShut {NoStop}%
\bibitem [{\citenamefont {Gruson}\ \emph {et~al.}(2016)\citenamefont {Gruson},
  \citenamefont {Barreau}, \citenamefont {{Jim{\'e}nez-Galan}}, \citenamefont
  {Risoud}, \citenamefont {Caillat}, \citenamefont {Maquet}, \citenamefont
  {Carr{\'e}}, \citenamefont {Lepetit}, \citenamefont {Hergott}, \citenamefont
  {Ruchon}, \citenamefont {Argenti}, \citenamefont {Ta{\"i}eb}, \citenamefont
  {Mart{\'i}n},\ and\ \citenamefont {Sali{\`e}res}}]{gruson2016}%
  \BibitemOpen
  \bibfield  {author} {\bibinfo {author} {\bibfnamefont {V.}~\bibnamefont
  {Gruson}}, \bibinfo {author} {\bibfnamefont {L.}~\bibnamefont {Barreau}},
  \bibinfo {author} {\bibfnamefont {{\'A}.}~\bibnamefont
  {{Jim{\'e}nez-Galan}}}, \bibinfo {author} {\bibfnamefont {F.}~\bibnamefont
  {Risoud}}, \bibinfo {author} {\bibfnamefont {J.}~\bibnamefont {Caillat}},
  \bibinfo {author} {\bibfnamefont {A.}~\bibnamefont {Maquet}}, \bibinfo
  {author} {\bibfnamefont {B.}~\bibnamefont {Carr{\'e}}}, \bibinfo {author}
  {\bibfnamefont {F.}~\bibnamefont {Lepetit}}, \bibinfo {author} {\bibfnamefont
  {J.-F.}\ \bibnamefont {Hergott}}, \bibinfo {author} {\bibfnamefont
  {T.}~\bibnamefont {Ruchon}}, \bibinfo {author} {\bibfnamefont
  {L.}~\bibnamefont {Argenti}}, \bibinfo {author} {\bibfnamefont
  {R.}~\bibnamefont {Ta{\"i}eb}}, \bibinfo {author} {\bibfnamefont
  {F.}~\bibnamefont {Mart{\'i}n}},\ and\ \bibinfo {author} {\bibfnamefont
  {P.}~\bibnamefont {Sali{\`e}res}},\ }\href
  {https://doi.org/10.1126/science.aah5188} {\bibfield  {journal} {\bibinfo
  {journal} {Science}\ }\textbf {\bibinfo {volume} {354}},\ \bibinfo {pages}
  {734} (\bibinfo {year} {2016})}\BibitemShut {NoStop}%
\bibitem [{\citenamefont {Sansone}\ \emph {et~al.}(2010)\citenamefont
  {Sansone}, \citenamefont {Kelkensberg}, \citenamefont {{P{\'e}rez-Torres}},
  \citenamefont {Morales}, \citenamefont {Kling}, \citenamefont {Siu},
  \citenamefont {Ghafur}, \citenamefont {Johnsson}, \citenamefont {Swoboda},
  \citenamefont {Benedetti}, \citenamefont {Ferrari}, \citenamefont
  {L{\'e}pine}, \citenamefont {{Sanz-Vicario}}, \citenamefont {Zherebtsov},
  \citenamefont {Znakovskaya}, \citenamefont {L'Huillier}, \citenamefont
  {Ivanov}, \citenamefont {Nisoli}, \citenamefont {Mart{\'i}n},\ and\
  \citenamefont {Vrakking}}]{sansone2010}%
  \BibitemOpen
  \bibfield  {author} {\bibinfo {author} {\bibfnamefont {G.}~\bibnamefont
  {Sansone}}, \bibinfo {author} {\bibfnamefont {F.}~\bibnamefont
  {Kelkensberg}}, \bibinfo {author} {\bibfnamefont {J.~F.}\ \bibnamefont
  {{P{\'e}rez-Torres}}}, \bibinfo {author} {\bibfnamefont {F.}~\bibnamefont
  {Morales}}, \bibinfo {author} {\bibfnamefont {M.~F.}\ \bibnamefont {Kling}},
  \bibinfo {author} {\bibfnamefont {W.}~\bibnamefont {Siu}}, \bibinfo {author}
  {\bibfnamefont {O.}~\bibnamefont {Ghafur}}, \bibinfo {author} {\bibfnamefont
  {P.}~\bibnamefont {Johnsson}}, \bibinfo {author} {\bibfnamefont
  {M.}~\bibnamefont {Swoboda}}, \bibinfo {author} {\bibfnamefont
  {E.}~\bibnamefont {Benedetti}}, \bibinfo {author} {\bibfnamefont
  {F.}~\bibnamefont {Ferrari}}, \bibinfo {author} {\bibfnamefont
  {F.}~\bibnamefont {L{\'e}pine}}, \bibinfo {author} {\bibfnamefont {J.~L.}\
  \bibnamefont {{Sanz-Vicario}}}, \bibinfo {author} {\bibfnamefont
  {S.}~\bibnamefont {Zherebtsov}}, \bibinfo {author} {\bibfnamefont
  {I.}~\bibnamefont {Znakovskaya}}, \bibinfo {author} {\bibfnamefont
  {A.}~\bibnamefont {L'Huillier}}, \bibinfo {author} {\bibfnamefont {M.~{\relax
  Yu}.}\ \bibnamefont {Ivanov}}, \bibinfo {author} {\bibfnamefont
  {M.}~\bibnamefont {Nisoli}}, \bibinfo {author} {\bibfnamefont
  {F.}~\bibnamefont {Mart{\'i}n}},\ and\ \bibinfo {author} {\bibfnamefont
  {M.~J.~J.}\ \bibnamefont {Vrakking}},\ }\href
  {https://doi.org/10.1038/nature09084} {\bibfield  {journal} {\bibinfo
  {journal} {Nature}\ }\textbf {\bibinfo {volume} {465}},\ \bibinfo {pages}
  {763} (\bibinfo {year} {2010})}\BibitemShut {NoStop}%
\bibitem [{\citenamefont {Cowley}(1995)}]{cowley1995}%
  \BibitemOpen
  \bibfield  {author} {\bibinfo {author} {\bibfnamefont {J.~M.}\ \bibnamefont
  {Cowley}},\ }\href@noop {} {\emph {\bibinfo {title} {Diffraction Physics}}},\
  \bibinfo {edition} {3rd}\ ed.,\ North-{{Holland}} Personal Library\ (\bibinfo
   {publisher} {Elsevier Science B.V},\ \bibinfo {address} {Amsterdam ; New
  York},\ \bibinfo {year} {1995})\BibitemShut {NoStop}%
\bibitem [{\citenamefont {Schlichting}\ and\ \citenamefont
  {Miao}(2012)}]{schlichting2012}%
  \BibitemOpen
  \bibfield  {author} {\bibinfo {author} {\bibfnamefont {I.}~\bibnamefont
  {Schlichting}}\ and\ \bibinfo {author} {\bibfnamefont {J.}~\bibnamefont
  {Miao}},\ }\href {https://doi.org/10.1016/j.sbi.2012.07.015} {\bibfield
  {journal} {\bibinfo  {journal} {Curr. Opin. Struc. Biol.}\ }\textbf {\bibinfo
  {volume} {22}},\ \bibinfo {pages} {613} (\bibinfo {year} {2012})}\BibitemShut
  {NoStop}%
\bibitem [{\citenamefont {Colliex}\ \emph {et~al.}(2006)\citenamefont
  {Colliex}, \citenamefont {Cowley}, \citenamefont {Dudarev}, \citenamefont
  {Fink}, \citenamefont {Gj{\o}nnes}, \citenamefont {Hilderbrandt},
  \citenamefont {Howie}, \citenamefont {Lynch}, \citenamefont {Peng},
  \citenamefont {Ren}, \citenamefont {Ross}, \citenamefont {Smith},
  \citenamefont {Spence}, \citenamefont {Steeds}, \citenamefont {Wang},
  \citenamefont {Whelan},\ and\ \citenamefont {Zvyagin}}]{colliex2006}%
  \BibitemOpen
  \bibfield  {author} {\bibinfo {author} {\bibfnamefont {C.}~\bibnamefont
  {Colliex}}, \bibinfo {author} {\bibfnamefont {J.~M.}\ \bibnamefont {Cowley}},
  \bibinfo {author} {\bibfnamefont {S.~L.}\ \bibnamefont {Dudarev}}, \bibinfo
  {author} {\bibfnamefont {M.}~\bibnamefont {Fink}}, \bibinfo {author}
  {\bibfnamefont {J.}~\bibnamefont {Gj{\o}nnes}}, \bibinfo {author}
  {\bibfnamefont {R.}~\bibnamefont {Hilderbrandt}}, \bibinfo {author}
  {\bibfnamefont {A.}~\bibnamefont {Howie}}, \bibinfo {author} {\bibfnamefont
  {D.~F.}\ \bibnamefont {Lynch}}, \bibinfo {author} {\bibfnamefont {L.~M.}\
  \bibnamefont {Peng}}, \bibinfo {author} {\bibfnamefont {G.}~\bibnamefont
  {Ren}}, \bibinfo {author} {\bibfnamefont {A.~W.}\ \bibnamefont {Ross}},
  \bibinfo {author} {\bibfnamefont {V.~H.}\ \bibnamefont {Smith}}, \bibinfo
  {author} {\bibfnamefont {J.~C.~H.}\ \bibnamefont {Spence}}, \bibinfo {author}
  {\bibfnamefont {J.~W.}\ \bibnamefont {Steeds}}, \bibinfo {author}
  {\bibfnamefont {J.}~\bibnamefont {Wang}}, \bibinfo {author} {\bibfnamefont
  {M.~J.}\ \bibnamefont {Whelan}},\ and\ \bibinfo {author} {\bibfnamefont
  {B.~B.}\ \bibnamefont {Zvyagin}},\ }in\ \href
  {https://doi.org/10.1107/97809553602060000593} {\emph {\bibinfo {booktitle}
  {International {{Tables}} for {{Crystallography}}}}},\ Vol.~\bibinfo {volume}
  {C},\ \bibinfo {editor} {edited by\ \bibinfo {editor} {\bibfnamefont
  {H.}~\bibnamefont {Fuess}}, \bibinfo {editor} {\bibfnamefont {{\relax
  Th}.}~\bibnamefont {Hahn}}, \bibinfo {editor} {\bibfnamefont
  {H.}~\bibnamefont {Wondratschek}}, \bibinfo {editor} {\bibfnamefont
  {U.}~\bibnamefont {M{\"u}ller}}, \bibinfo {editor} {\bibfnamefont
  {U.}~\bibnamefont {Shmueli}}, \bibinfo {editor} {\bibfnamefont
  {E.}~\bibnamefont {Prince}}, \bibinfo {editor} {\bibfnamefont
  {A.}~\bibnamefont {Authier}}, \bibinfo {editor} {\bibfnamefont
  {V.}~\bibnamefont {Kopsk{\'y}}}, \bibinfo {editor} {\bibfnamefont {D.~B.}\
  \bibnamefont {Litvin}}, \bibinfo {editor} {\bibfnamefont {M.~G.}\
  \bibnamefont {Rossmann}}, \bibinfo {editor} {\bibfnamefont {E.}~\bibnamefont
  {Arnold}}, \bibinfo {editor} {\bibfnamefont {S.}~\bibnamefont {Hall}},
  \bibinfo {editor} {\bibfnamefont {B.}~\bibnamefont {McMahon}},\ and\ \bibinfo
  {editor} {\bibfnamefont {E.}~\bibnamefont {Prince}}}\ (\bibinfo  {publisher}
  {International Union of Crystallography},\ \bibinfo {address} {Chester,
  England},\ \bibinfo {year} {2006})\ \bibinfo {edition} {1st}\ ed.,\ pp.\
  \bibinfo {pages} {259--429}\BibitemShut {NoStop}%
\bibitem [{\citenamefont {Carter}(2016)}]{carter2016}%
  \BibitemOpen
  \bibfield  {author} {\bibinfo {author} {\bibfnamefont {C.~B.}\ \bibnamefont
  {Carter}},\ }\href@noop {} {\emph {\bibinfo {title} {Transmission Electron
  Microscopy: Diffraction, Imaging, and Spectrometry}}},\ \bibinfo {edition}
  {1st}\ ed.\ (\bibinfo  {publisher} {Springer Berlin Heidelberg},\ \bibinfo
  {address} {New York, NY},\ \bibinfo {year} {2016})\BibitemShut {NoStop}%
\bibitem [{\citenamefont {Hassan}\ \emph {et~al.}(2017)\citenamefont {Hassan},
  \citenamefont {Baskin}, \citenamefont {Liao},\ and\ \citenamefont
  {Zewail}}]{Hassan2017}%
  \BibitemOpen
  \bibfield  {author} {\bibinfo {author} {\bibfnamefont {M.~{\relax Th}.}\
  \bibnamefont {Hassan}}, \bibinfo {author} {\bibfnamefont {J.~S.}\
  \bibnamefont {Baskin}}, \bibinfo {author} {\bibfnamefont {B.}~\bibnamefont
  {Liao}},\ and\ \bibinfo {author} {\bibfnamefont {A.~H.}\ \bibnamefont
  {Zewail}},\ }\href {https://doi.org/10.1038/nphoton.2017.79} {\bibfield
  {journal} {\bibinfo  {journal} {Nature Photonics}\ }\textbf {\bibinfo
  {volume} {11}},\ \bibinfo {pages} {425} (\bibinfo {year} {2017})}\BibitemShut
  {NoStop}%
\bibitem [{\citenamefont {Hassan}(2018)}]{Hassan2018}%
  \BibitemOpen
  \bibfield  {author} {\bibinfo {author} {\bibfnamefont {M.~T.}\ \bibnamefont
  {Hassan}},\ }\href {https://doi.org/10.1088/1361-6455/aaa183} {\bibfield
  {journal} {\bibinfo  {journal} {Journal of Physics B: Atomic, Molecular and
  Optical Physics}\ }\textbf {\bibinfo {volume} {51}},\ \bibinfo {pages}
  {032005} (\bibinfo {year} {2018})}\BibitemShut {NoStop}%
\bibitem [{\citenamefont {Ishikawa}\ \emph {et~al.}(2012)\citenamefont
  {Ishikawa}, \citenamefont {Aoyagi}, \citenamefont {Asaka}, \citenamefont
  {Asano}, \citenamefont {Azumi}, \citenamefont {Bizen}, \citenamefont {Ego},
  \citenamefont {Fukami}, \citenamefont {Fukui}, \citenamefont {Furukawa},
  \citenamefont {Goto}, \citenamefont {Hanaki}, \citenamefont {Hara},
  \citenamefont {Hasegawa}, \citenamefont {Hatsui}, \citenamefont {Higashiya},
  \citenamefont {Hirono}, \citenamefont {Hosoda}, \citenamefont {Ishii},
  \citenamefont {Inagaki}, \citenamefont {Inubushi}, \citenamefont {Itoga},
  \citenamefont {Joti}, \citenamefont {Kago}, \citenamefont {Kameshima},
  \citenamefont {Kimura}, \citenamefont {Kirihara}, \citenamefont {Kiyomichi},
  \citenamefont {Kobayashi}, \citenamefont {Kondo}, \citenamefont {Kudo},
  \citenamefont {Maesaka}, \citenamefont {Mar{\'e}chal}, \citenamefont
  {Masuda}, \citenamefont {Matsubara}, \citenamefont {Matsumoto}, \citenamefont
  {Matsushita}, \citenamefont {Matsui}, \citenamefont {Nagasono}, \citenamefont
  {Nariyama}, \citenamefont {Ohashi}, \citenamefont {Ohata}, \citenamefont
  {Ohshima}, \citenamefont {Ono}, \citenamefont {Otake}, \citenamefont {Saji},
  \citenamefont {Sakurai}, \citenamefont {Sato}, \citenamefont {Sawada},
  \citenamefont {Seike}, \citenamefont {Shirasawa}, \citenamefont {Sugimoto},
  \citenamefont {Suzuki}, \citenamefont {Takahashi}, \citenamefont {Takebe},
  \citenamefont {Takeshita}, \citenamefont {Tamasaku}, \citenamefont {Tanaka},
  \citenamefont {Tanaka}, \citenamefont {Tanaka}, \citenamefont {Togashi},
  \citenamefont {Togawa}, \citenamefont {Tokuhisa}, \citenamefont {Tomizawa},
  \citenamefont {Tono}, \citenamefont {Wu}, \citenamefont {Yabashi},
  \citenamefont {Yamaga}, \citenamefont {Yamashita}, \citenamefont {Yanagida},
  \citenamefont {Zhang}, \citenamefont {Shintake}, \citenamefont {Kitamura},\
  and\ \citenamefont {Kumagai}}]{ishikawa2012}%
  \BibitemOpen
  \bibfield  {author} {\bibinfo {author} {\bibfnamefont {T.}~\bibnamefont
  {Ishikawa}}, \bibinfo {author} {\bibfnamefont {H.}~\bibnamefont {Aoyagi}},
  \bibinfo {author} {\bibfnamefont {T.}~\bibnamefont {Asaka}}, \bibinfo
  {author} {\bibfnamefont {Y.}~\bibnamefont {Asano}}, \bibinfo {author}
  {\bibfnamefont {N.}~\bibnamefont {Azumi}}, \bibinfo {author} {\bibfnamefont
  {T.}~\bibnamefont {Bizen}}, \bibinfo {author} {\bibfnamefont
  {H.}~\bibnamefont {Ego}}, \bibinfo {author} {\bibfnamefont {K.}~\bibnamefont
  {Fukami}}, \bibinfo {author} {\bibfnamefont {T.}~\bibnamefont {Fukui}},
  \bibinfo {author} {\bibfnamefont {Y.}~\bibnamefont {Furukawa}}, \bibinfo
  {author} {\bibfnamefont {S.}~\bibnamefont {Goto}}, \bibinfo {author}
  {\bibfnamefont {H.}~\bibnamefont {Hanaki}}, \bibinfo {author} {\bibfnamefont
  {T.}~\bibnamefont {Hara}}, \bibinfo {author} {\bibfnamefont {T.}~\bibnamefont
  {Hasegawa}}, \bibinfo {author} {\bibfnamefont {T.}~\bibnamefont {Hatsui}},
  \bibinfo {author} {\bibfnamefont {A.}~\bibnamefont {Higashiya}}, \bibinfo
  {author} {\bibfnamefont {T.}~\bibnamefont {Hirono}}, \bibinfo {author}
  {\bibfnamefont {N.}~\bibnamefont {Hosoda}}, \bibinfo {author} {\bibfnamefont
  {M.}~\bibnamefont {Ishii}}, \bibinfo {author} {\bibfnamefont
  {T.}~\bibnamefont {Inagaki}}, \bibinfo {author} {\bibfnamefont
  {Y.}~\bibnamefont {Inubushi}}, \bibinfo {author} {\bibfnamefont
  {T.}~\bibnamefont {Itoga}}, \bibinfo {author} {\bibfnamefont
  {Y.}~\bibnamefont {Joti}}, \bibinfo {author} {\bibfnamefont {M.}~\bibnamefont
  {Kago}}, \bibinfo {author} {\bibfnamefont {T.}~\bibnamefont {Kameshima}},
  \bibinfo {author} {\bibfnamefont {H.}~\bibnamefont {Kimura}}, \bibinfo
  {author} {\bibfnamefont {Y.}~\bibnamefont {Kirihara}}, \bibinfo {author}
  {\bibfnamefont {A.}~\bibnamefont {Kiyomichi}}, \bibinfo {author}
  {\bibfnamefont {T.}~\bibnamefont {Kobayashi}}, \bibinfo {author}
  {\bibfnamefont {C.}~\bibnamefont {Kondo}}, \bibinfo {author} {\bibfnamefont
  {T.}~\bibnamefont {Kudo}}, \bibinfo {author} {\bibfnamefont {H.}~\bibnamefont
  {Maesaka}}, \bibinfo {author} {\bibfnamefont {X.~M.}\ \bibnamefont
  {Mar{\'e}chal}}, \bibinfo {author} {\bibfnamefont {T.}~\bibnamefont
  {Masuda}}, \bibinfo {author} {\bibfnamefont {S.}~\bibnamefont {Matsubara}},
  \bibinfo {author} {\bibfnamefont {T.}~\bibnamefont {Matsumoto}}, \bibinfo
  {author} {\bibfnamefont {T.}~\bibnamefont {Matsushita}}, \bibinfo {author}
  {\bibfnamefont {S.}~\bibnamefont {Matsui}}, \bibinfo {author} {\bibfnamefont
  {M.}~\bibnamefont {Nagasono}}, \bibinfo {author} {\bibfnamefont
  {N.}~\bibnamefont {Nariyama}}, \bibinfo {author} {\bibfnamefont
  {H.}~\bibnamefont {Ohashi}}, \bibinfo {author} {\bibfnamefont
  {T.}~\bibnamefont {Ohata}}, \bibinfo {author} {\bibfnamefont
  {T.}~\bibnamefont {Ohshima}}, \bibinfo {author} {\bibfnamefont
  {S.}~\bibnamefont {Ono}}, \bibinfo {author} {\bibfnamefont {Y.}~\bibnamefont
  {Otake}}, \bibinfo {author} {\bibfnamefont {C.}~\bibnamefont {Saji}},
  \bibinfo {author} {\bibfnamefont {T.}~\bibnamefont {Sakurai}}, \bibinfo
  {author} {\bibfnamefont {T.}~\bibnamefont {Sato}}, \bibinfo {author}
  {\bibfnamefont {K.}~\bibnamefont {Sawada}}, \bibinfo {author} {\bibfnamefont
  {T.}~\bibnamefont {Seike}}, \bibinfo {author} {\bibfnamefont
  {K.}~\bibnamefont {Shirasawa}}, \bibinfo {author} {\bibfnamefont
  {T.}~\bibnamefont {Sugimoto}}, \bibinfo {author} {\bibfnamefont
  {S.}~\bibnamefont {Suzuki}}, \bibinfo {author} {\bibfnamefont
  {S.}~\bibnamefont {Takahashi}}, \bibinfo {author} {\bibfnamefont
  {H.}~\bibnamefont {Takebe}}, \bibinfo {author} {\bibfnamefont
  {K.}~\bibnamefont {Takeshita}}, \bibinfo {author} {\bibfnamefont
  {K.}~\bibnamefont {Tamasaku}}, \bibinfo {author} {\bibfnamefont
  {H.}~\bibnamefont {Tanaka}}, \bibinfo {author} {\bibfnamefont
  {R.}~\bibnamefont {Tanaka}}, \bibinfo {author} {\bibfnamefont
  {T.}~\bibnamefont {Tanaka}}, \bibinfo {author} {\bibfnamefont
  {T.}~\bibnamefont {Togashi}}, \bibinfo {author} {\bibfnamefont
  {K.}~\bibnamefont {Togawa}}, \bibinfo {author} {\bibfnamefont
  {A.}~\bibnamefont {Tokuhisa}}, \bibinfo {author} {\bibfnamefont
  {H.}~\bibnamefont {Tomizawa}}, \bibinfo {author} {\bibfnamefont
  {K.}~\bibnamefont {Tono}}, \bibinfo {author} {\bibfnamefont {S.}~\bibnamefont
  {Wu}}, \bibinfo {author} {\bibfnamefont {M.}~\bibnamefont {Yabashi}},
  \bibinfo {author} {\bibfnamefont {M.}~\bibnamefont {Yamaga}}, \bibinfo
  {author} {\bibfnamefont {A.}~\bibnamefont {Yamashita}}, \bibinfo {author}
  {\bibfnamefont {K.}~\bibnamefont {Yanagida}}, \bibinfo {author}
  {\bibfnamefont {C.}~\bibnamefont {Zhang}}, \bibinfo {author} {\bibfnamefont
  {T.}~\bibnamefont {Shintake}}, \bibinfo {author} {\bibfnamefont
  {H.}~\bibnamefont {Kitamura}},\ and\ \bibinfo {author} {\bibfnamefont
  {N.}~\bibnamefont {Kumagai}},\ }\href
  {https://doi.org/10.1038/nphoton.2012.141} {\bibfield  {journal} {\bibinfo
  {journal} {Nat. Photonics}\ }\textbf {\bibinfo {volume} {6}},\ \bibinfo
  {pages} {540} (\bibinfo {year} {2012})}\BibitemShut {NoStop}%
\bibitem [{\citenamefont {Pellegrini}\ \emph {et~al.}(2016)\citenamefont
  {Pellegrini}, \citenamefont {Marinelli},\ and\ \citenamefont
  {Reiche}}]{pellegrini2016}%
  \BibitemOpen
  \bibfield  {author} {\bibinfo {author} {\bibfnamefont {C.}~\bibnamefont
  {Pellegrini}}, \bibinfo {author} {\bibfnamefont {A.}~\bibnamefont
  {Marinelli}},\ and\ \bibinfo {author} {\bibfnamefont {S.}~\bibnamefont
  {Reiche}},\ }\href {https://doi.org/10.1103/RevModPhys.88.015006} {\bibfield
  {journal} {\bibinfo  {journal} {Rev. Mod. Phys.}\ }\textbf {\bibinfo {volume}
  {88}},\ \bibinfo {pages} {015006} (\bibinfo {year} {2016})}\BibitemShut
  {NoStop}%
\bibitem [{\citenamefont {Hui}\ \emph {et~al.}(2024)\citenamefont {Hui},
  \citenamefont {Alqattan}, \citenamefont {Sennary}, \citenamefont {Golubev},\
  and\ \citenamefont {Hassan}}]{hui2023}%
  \BibitemOpen
  \bibfield  {author} {\bibinfo {author} {\bibfnamefont {D.}~\bibnamefont
  {Hui}}, \bibinfo {author} {\bibfnamefont {H.}~\bibnamefont {Alqattan}},
  \bibinfo {author} {\bibfnamefont {M.}~\bibnamefont {Sennary}}, \bibinfo
  {author} {\bibfnamefont {N.~V.}\ \bibnamefont {Golubev}},\ and\ \bibinfo
  {author} {\bibfnamefont {M.~T.}\ \bibnamefont {Hassan}},\ }\href
  {https://doi.org/10.1126/sciadv.adp5805} {\bibfield  {journal} {\bibinfo
  {journal} {Sci. Adv.}\ }\textbf {\bibinfo {volume} {10}},\ \bibinfo {pages}
  {eadp5805} (\bibinfo {year} {2024})}\BibitemShut {NoStop}%
\bibitem [{\citenamefont {Coppens}(1992)}]{coppens1992}%
  \BibitemOpen
  \bibfield  {author} {\bibinfo {author} {\bibfnamefont {P.}~\bibnamefont
  {Coppens}},\ }\href {https://doi.org/10.1146/annurev.pc.43.100192.003311}
  {\bibfield  {journal} {\bibinfo  {journal} {Ann. Phys. Chem.}\ }\textbf
  {\bibinfo {volume} {43}},\ \bibinfo {pages} {663} (\bibinfo {year}
  {1992})}\BibitemShut {NoStop}%
\bibitem [{\citenamefont {Coppens}(1997)}]{coppens1997}%
  \BibitemOpen
  \bibfield  {author} {\bibinfo {author} {\bibfnamefont {P.}~\bibnamefont
  {Coppens}},\ }\href@noop {} {\emph {\bibinfo {title} {X-Ray Charge Densities
  and Chemical Bonding}}},\ \bibinfo {series} {International {{Union}} of
  {{Crystallography}} Texts on Crystallography}\ No.~\bibinfo {number} {4}\
  (\bibinfo  {publisher} {International Union of Crystallography ; Oxford
  University Press},\ \bibinfo {address} {[Chester, England] : Oxford ; New
  York},\ \bibinfo {year} {1997})\BibitemShut {NoStop}%
\bibitem [{\citenamefont {Cao}\ and\ \citenamefont {Wilson}(1998)}]{cao1998a}%
  \BibitemOpen
  \bibfield  {author} {\bibinfo {author} {\bibfnamefont {J.}~\bibnamefont
  {Cao}}\ and\ \bibinfo {author} {\bibfnamefont {K.~R.}\ \bibnamefont
  {Wilson}},\ }\href {https://doi.org/10.1021/jp982054p} {\bibfield  {journal}
  {\bibinfo  {journal} {J. Phys. Chem. A}\ }\textbf {\bibinfo {volume} {102}},\
  \bibinfo {pages} {9523} (\bibinfo {year} {1998})}\BibitemShut {NoStop}%
\bibitem [{\citenamefont {Authier}(2008)}]{authier2008}%
  \BibitemOpen
  \bibfield  {author} {\bibinfo {author} {\bibfnamefont {A.}~\bibnamefont
  {Authier}},\ }\href@noop {} {\emph {\bibinfo {title} {Dynamical Theory of
  {{X-ray}} Diffraction}}},\ \bibinfo {edition} {repr}\ ed.,\ \bibinfo {series}
  {Monographs on Crystallography}\ No.~\bibinfo {number} {11}\ (\bibinfo
  {publisher} {Oxford Univ. Press},\ \bibinfo {address} {Oxford},\ \bibinfo
  {year} {2008})\BibitemShut {NoStop}%
\bibitem [{\citenamefont {Centurion}\ \emph {et~al.}(2022)\citenamefont
  {Centurion}, \citenamefont {Wolf},\ and\ \citenamefont
  {Yang}}]{centurion2022}%
  \BibitemOpen
  \bibfield  {author} {\bibinfo {author} {\bibfnamefont {M.}~\bibnamefont
  {Centurion}}, \bibinfo {author} {\bibfnamefont {T.~J.}\ \bibnamefont
  {Wolf}},\ and\ \bibinfo {author} {\bibfnamefont {J.}~\bibnamefont {Yang}},\
  }\href {https://doi.org/10.1146/annurev-physchem-082720-010539} {\bibfield
  {journal} {\bibinfo  {journal} {Ann. Phys. Chem.}\ }\textbf {\bibinfo
  {volume} {73}},\ \bibinfo {pages} {21} (\bibinfo {year} {2022})}\BibitemShut
  {NoStop}%
\bibitem [{\citenamefont {Shao}\ and\ \citenamefont
  {Starace}(2010)}]{shao2010}%
  \BibitemOpen
  \bibfield  {author} {\bibinfo {author} {\bibfnamefont {H.-C.}\ \bibnamefont
  {Shao}}\ and\ \bibinfo {author} {\bibfnamefont {A.~F.}\ \bibnamefont
  {Starace}},\ }\href {https://doi.org/10.1103/PhysRevLett.105.263201}
  {\bibfield  {journal} {\bibinfo  {journal} {Phys. Rev. Lett.}\ }\textbf
  {\bibinfo {volume} {105}},\ \bibinfo {pages} {263201} (\bibinfo {year}
  {2010})}\BibitemShut {NoStop}%
\bibitem [{\citenamefont {Suominen}\ and\ \citenamefont
  {Kirrander}(2014)}]{suominen2014}%
  \BibitemOpen
  \bibfield  {author} {\bibinfo {author} {\bibfnamefont {H.~J.}\ \bibnamefont
  {Suominen}}\ and\ \bibinfo {author} {\bibfnamefont {A.}~\bibnamefont
  {Kirrander}},\ }\href {https://doi.org/10.1103/PhysRevLett.112.043002}
  {\bibfield  {journal} {\bibinfo  {journal} {Phys. Rev. Lett.}\ }\textbf
  {\bibinfo {volume} {112}},\ \bibinfo {pages} {043002} (\bibinfo {year}
  {2014})}\BibitemShut {NoStop}%
\bibitem [{\citenamefont {Baum}\ \emph {et~al.}(2010)\citenamefont {Baum},
  \citenamefont {Manz},\ and\ \citenamefont {Schild}}]{baum2010}%
  \BibitemOpen
  \bibfield  {author} {\bibinfo {author} {\bibfnamefont {P.}~\bibnamefont
  {Baum}}, \bibinfo {author} {\bibfnamefont {J.}~\bibnamefont {Manz}},\ and\
  \bibinfo {author} {\bibfnamefont {A.}~\bibnamefont {Schild}},\ }\href
  {https://doi.org/10.1007/s11433-010-4017-y} {\bibfield  {journal} {\bibinfo
  {journal} {Sci. China Phys. Mech. Astron.}\ }\textbf {\bibinfo {volume}
  {53}},\ \bibinfo {pages} {987} (\bibinfo {year} {2010})}\BibitemShut
  {NoStop}%
\bibitem [{\citenamefont {Yakovlev}\ \emph {et~al.}(2015)\citenamefont
  {Yakovlev}, \citenamefont {Stockman}, \citenamefont {Krausz},\ and\
  \citenamefont {Baum}}]{yakovlev2015}%
  \BibitemOpen
  \bibfield  {author} {\bibinfo {author} {\bibfnamefont {V.~S.}\ \bibnamefont
  {Yakovlev}}, \bibinfo {author} {\bibfnamefont {M.~I.}\ \bibnamefont
  {Stockman}}, \bibinfo {author} {\bibfnamefont {F.}~\bibnamefont {Krausz}},\
  and\ \bibinfo {author} {\bibfnamefont {P.}~\bibnamefont {Baum}},\ }\href
  {https://doi.org/10.1038/srep14581} {\bibfield  {journal} {\bibinfo
  {journal} {Sci. Rep.}\ }\textbf {\bibinfo {volume} {5}},\ \bibinfo {pages}
  {14581} (\bibinfo {year} {2015})}\BibitemShut {NoStop}%
\bibitem [{\citenamefont {Dixit}\ \emph {et~al.}(2012)\citenamefont {Dixit},
  \citenamefont {Vendrell},\ and\ \citenamefont {Santra}}]{dixit2012}%
  \BibitemOpen
  \bibfield  {author} {\bibinfo {author} {\bibfnamefont {G.}~\bibnamefont
  {Dixit}}, \bibinfo {author} {\bibfnamefont {O.}~\bibnamefont {Vendrell}},\
  and\ \bibinfo {author} {\bibfnamefont {R.}~\bibnamefont {Santra}},\ }\href
  {https://doi.org/10.1073/pnas.1202226109} {\bibfield  {journal} {\bibinfo
  {journal} {Proc. Natl. Acad. Sci.}\ }\textbf {\bibinfo {volume} {109}},\
  \bibinfo {pages} {11636} (\bibinfo {year} {2012})}\BibitemShut {NoStop}%
\bibitem [{\citenamefont {Dixit}\ \emph {et~al.}(2014)\citenamefont {Dixit},
  \citenamefont {Slowik},\ and\ \citenamefont {Santra}}]{dixit2014}%
  \BibitemOpen
  \bibfield  {author} {\bibinfo {author} {\bibfnamefont {G.}~\bibnamefont
  {Dixit}}, \bibinfo {author} {\bibfnamefont {J.~M.}\ \bibnamefont {Slowik}},\
  and\ \bibinfo {author} {\bibfnamefont {R.}~\bibnamefont {Santra}},\ }\href
  {https://doi.org/10.1103/PhysRevA.89.043409} {\bibfield  {journal} {\bibinfo
  {journal} {Phys. Rev. A}\ }\textbf {\bibinfo {volume} {89}},\ \bibinfo
  {pages} {043409} (\bibinfo {year} {2014})}\BibitemShut {NoStop}%
\bibitem [{\citenamefont {Dixit}\ \emph {et~al.}(2013)\citenamefont {Dixit},
  \citenamefont {Slowik},\ and\ \citenamefont {Santra}}]{dixit2013}%
  \BibitemOpen
  \bibfield  {author} {\bibinfo {author} {\bibfnamefont {G.}~\bibnamefont
  {Dixit}}, \bibinfo {author} {\bibfnamefont {J.~M.}\ \bibnamefont {Slowik}},\
  and\ \bibinfo {author} {\bibfnamefont {R.}~\bibnamefont {Santra}},\ }\href
  {https://doi.org/10.1103/PhysRevLett.110.137403} {\bibfield  {journal}
  {\bibinfo  {journal} {Phys. Rev. Lett.}\ }\textbf {\bibinfo {volume} {110}},\
  \bibinfo {pages} {137403} (\bibinfo {year} {2013})}\BibitemShut {NoStop}%
\bibitem [{\citenamefont {Dixit}\ and\ \citenamefont
  {Santra}(2013)}]{dixit2013a}%
  \BibitemOpen
  \bibfield  {author} {\bibinfo {author} {\bibfnamefont {G.}~\bibnamefont
  {Dixit}}\ and\ \bibinfo {author} {\bibfnamefont {R.}~\bibnamefont {Santra}},\
  }\href {https://doi.org/10.1063/1.4798321} {\bibfield  {journal} {\bibinfo
  {journal} {J. Chem. Phys.}\ }\textbf {\bibinfo {volume} {138}},\ \bibinfo
  {pages} {134311} (\bibinfo {year} {2013})}\BibitemShut {NoStop}%
\bibitem [{\citenamefont {Shao}\ and\ \citenamefont
  {Starace}(2013)}]{shao2013}%
  \BibitemOpen
  \bibfield  {author} {\bibinfo {author} {\bibfnamefont {H.-C.}\ \bibnamefont
  {Shao}}\ and\ \bibinfo {author} {\bibfnamefont {A.~F.}\ \bibnamefont
  {Starace}},\ }\href {https://doi.org/10.1103/PhysRevA.88.062711} {\bibfield
  {journal} {\bibinfo  {journal} {Phys. Rev. A}\ }\textbf {\bibinfo {volume}
  {88}},\ \bibinfo {pages} {062711} (\bibinfo {year} {2013})}\BibitemShut
  {NoStop}%
\bibitem [{\citenamefont {Shao}\ and\ \citenamefont
  {Starace}(2014)}]{shao2014}%
  \BibitemOpen
  \bibfield  {author} {\bibinfo {author} {\bibfnamefont {H.-C.}\ \bibnamefont
  {Shao}}\ and\ \bibinfo {author} {\bibfnamefont {A.~F.}\ \bibnamefont
  {Starace}},\ }\href {https://doi.org/10.1103/PhysRevA.90.032710} {\bibfield
  {journal} {\bibinfo  {journal} {Phys. Rev. A}\ }\textbf {\bibinfo {volume}
  {90}},\ \bibinfo {pages} {032710} (\bibinfo {year} {2014})}\BibitemShut
  {NoStop}%
\bibitem [{\citenamefont {Shao}\ and\ \citenamefont
  {Starace}(2016)}]{shao2016}%
  \BibitemOpen
  \bibfield  {author} {\bibinfo {author} {\bibfnamefont {H.-C.}\ \bibnamefont
  {Shao}}\ and\ \bibinfo {author} {\bibfnamefont {A.~F.}\ \bibnamefont
  {Starace}},\ }\href {https://doi.org/10.1103/PhysRevA.94.030702} {\bibfield
  {journal} {\bibinfo  {journal} {Phys. Rev. A}\ }\textbf {\bibinfo {volume}
  {94}},\ \bibinfo {pages} {030702} (\bibinfo {year} {2016})}\BibitemShut
  {NoStop}%
\bibitem [{\citenamefont {Dixit}\ and\ \citenamefont
  {Santra}(2017)}]{dixit2017}%
  \BibitemOpen
  \bibfield  {author} {\bibinfo {author} {\bibfnamefont {G.}~\bibnamefont
  {Dixit}}\ and\ \bibinfo {author} {\bibfnamefont {R.}~\bibnamefont {Santra}},\
  }\href {https://doi.org/10.1103/PhysRevA.96.053413} {\bibfield  {journal}
  {\bibinfo  {journal} {Phys. Rev. A}\ }\textbf {\bibinfo {volume} {96}},\
  \bibinfo {pages} {053413} (\bibinfo {year} {2017})}\BibitemShut {NoStop}%
\bibitem [{\citenamefont {Grosser}\ \emph {et~al.}(2017)\citenamefont
  {Grosser}, \citenamefont {Slowik},\ and\ \citenamefont
  {Santra}}]{grosser2017}%
  \BibitemOpen
  \bibfield  {author} {\bibinfo {author} {\bibfnamefont {M.}~\bibnamefont
  {Grosser}}, \bibinfo {author} {\bibfnamefont {J.~M.}\ \bibnamefont
  {Slowik}},\ and\ \bibinfo {author} {\bibfnamefont {R.}~\bibnamefont
  {Santra}},\ }\href {https://doi.org/10.1103/PhysRevA.95.062107} {\bibfield
  {journal} {\bibinfo  {journal} {Phys. Rev. A}\ }\textbf {\bibinfo {volume}
  {95}},\ \bibinfo {pages} {062107} (\bibinfo {year} {2017})}\BibitemShut
  {NoStop}%
\bibitem [{\citenamefont {Shao}\ and\ \citenamefont
  {Starace}(2017)}]{shao2017}%
  \BibitemOpen
  \bibfield  {author} {\bibinfo {author} {\bibfnamefont {H.-C.}\ \bibnamefont
  {Shao}}\ and\ \bibinfo {author} {\bibfnamefont {A.~F.}\ \bibnamefont
  {Starace}},\ }in\ \href {https://doi.org/10.1117/12.2273560} {\emph {\bibinfo
  {booktitle} {Ultrafast {{Nonlinear Imaging}} and {{Spectroscopy V}}}}},\
  \bibinfo {editor} {edited by\ \bibinfo {editor} {\bibfnamefont
  {Z.}~\bibnamefont {Liu}}}\ (\bibinfo  {publisher} {SPIE},\ \bibinfo {address}
  {San Diego, United States},\ \bibinfo {year} {2017})\ p.~\bibinfo {pages}
  {12}\BibitemShut {NoStop}%
\bibitem [{\citenamefont {Bennett}\ \emph {et~al.}(2014)\citenamefont
  {Bennett}, \citenamefont {Biggs}, \citenamefont {Zhang}, \citenamefont
  {Dorfman},\ and\ \citenamefont {Mukamel}}]{bennett2014}%
  \BibitemOpen
  \bibfield  {author} {\bibinfo {author} {\bibfnamefont {K.}~\bibnamefont
  {Bennett}}, \bibinfo {author} {\bibfnamefont {J.~D.}\ \bibnamefont {Biggs}},
  \bibinfo {author} {\bibfnamefont {Y.}~\bibnamefont {Zhang}}, \bibinfo
  {author} {\bibfnamefont {K.~E.}\ \bibnamefont {Dorfman}},\ and\ \bibinfo
  {author} {\bibfnamefont {S.}~\bibnamefont {Mukamel}},\ }\href
  {https://doi.org/10.1063/1.4878377} {\bibfield  {journal} {\bibinfo
  {journal} {J. Chem. Phys.}\ }\textbf {\bibinfo {volume} {140}},\ \bibinfo
  {pages} {204311} (\bibinfo {year} {2014})}\BibitemShut {NoStop}%
\bibitem [{\citenamefont {Hermann}\ \emph {et~al.}(2020)\citenamefont
  {Hermann}, \citenamefont {Pohl}, \citenamefont {Dixit},\ and\ \citenamefont
  {Tremblay}}]{hermann2020}%
  \BibitemOpen
  \bibfield  {author} {\bibinfo {author} {\bibfnamefont {G.}~\bibnamefont
  {Hermann}}, \bibinfo {author} {\bibfnamefont {V.}~\bibnamefont {Pohl}},
  \bibinfo {author} {\bibfnamefont {G.}~\bibnamefont {Dixit}},\ and\ \bibinfo
  {author} {\bibfnamefont {J.~C.}\ \bibnamefont {Tremblay}},\ }\href
  {https://doi.org/10.1103/PhysRevLett.124.013002} {\bibfield  {journal}
  {\bibinfo  {journal} {Phys. Rev. Lett.}\ }\textbf {\bibinfo {volume} {124}},\
  \bibinfo {pages} {013002} (\bibinfo {year} {2020})}\BibitemShut {NoStop}%
\bibitem [{\citenamefont {Giri}\ \emph {et~al.}(2021)\citenamefont {Giri},
  \citenamefont {Tremblay},\ and\ \citenamefont {Dixit}}]{giri2021}%
  \BibitemOpen
  \bibfield  {author} {\bibinfo {author} {\bibfnamefont {S.}~\bibnamefont
  {Giri}}, \bibinfo {author} {\bibfnamefont {J.~C.}\ \bibnamefont {Tremblay}},\
  and\ \bibinfo {author} {\bibfnamefont {G.}~\bibnamefont {Dixit}},\ }\href
  {https://doi.org/10.1103/PhysRevA.104.053115} {\bibfield  {journal} {\bibinfo
   {journal} {Phys. Rev. A}\ }\textbf {\bibinfo {volume} {104}},\ \bibinfo
  {pages} {053115} (\bibinfo {year} {2021})}\BibitemShut {NoStop}%
\bibitem [{\citenamefont {Rouxel}\ \emph {et~al.}(2021)\citenamefont {Rouxel},
  \citenamefont {Keefer},\ and\ \citenamefont {Mukamel}}]{rouxel2021}%
  \BibitemOpen
  \bibfield  {author} {\bibinfo {author} {\bibfnamefont {J.~R.}\ \bibnamefont
  {Rouxel}}, \bibinfo {author} {\bibfnamefont {D.}~\bibnamefont {Keefer}},\
  and\ \bibinfo {author} {\bibfnamefont {S.}~\bibnamefont {Mukamel}},\ }\href
  {https://doi.org/10.1063/4.0000043} {\bibfield  {journal} {\bibinfo
  {journal} {Struct. Dyn.}\ }\textbf {\bibinfo {volume} {8}},\ \bibinfo {pages}
  {014101} (\bibinfo {year} {2021})}\BibitemShut {NoStop}%
\bibitem [{\citenamefont {Tremblay}\ \emph {et~al.}(2021)\citenamefont
  {Tremblay}, \citenamefont {Pohl}, \citenamefont {Hermann},\ and\
  \citenamefont {Dixit}}]{tremblay2021}%
  \BibitemOpen
  \bibfield  {author} {\bibinfo {author} {\bibfnamefont {J.~C.}\ \bibnamefont
  {Tremblay}}, \bibinfo {author} {\bibfnamefont {V.}~\bibnamefont {Pohl}},
  \bibinfo {author} {\bibfnamefont {G.}~\bibnamefont {Hermann}},\ and\ \bibinfo
  {author} {\bibfnamefont {G.}~\bibnamefont {Dixit}},\ }\href
  {https://doi.org/10.1039/D0FD00116C} {\bibfield  {journal} {\bibinfo
  {journal} {Faraday Discuss.}\ }\textbf {\bibinfo {volume} {228}},\ \bibinfo
  {pages} {82} (\bibinfo {year} {2021})}\BibitemShut {NoStop}%
\bibitem [{\citenamefont {Kowalewski}\ \emph {et~al.}(2017)\citenamefont
  {Kowalewski}, \citenamefont {Bennett},\ and\ \citenamefont
  {Mukamel}}]{kowalewski2017}%
  \BibitemOpen
  \bibfield  {author} {\bibinfo {author} {\bibfnamefont {M.}~\bibnamefont
  {Kowalewski}}, \bibinfo {author} {\bibfnamefont {K.}~\bibnamefont
  {Bennett}},\ and\ \bibinfo {author} {\bibfnamefont {S.}~\bibnamefont
  {Mukamel}},\ }\href {https://doi.org/10.1063/1.4984241} {\bibfield  {journal}
  {\bibinfo  {journal} {Struct. Dyn.}\ }\textbf {\bibinfo {volume} {4}},\
  \bibinfo {pages} {054101} (\bibinfo {year} {2017})}\BibitemShut {NoStop}%
\bibitem [{\citenamefont {Bennett}\ \emph {et~al.}(2018)\citenamefont
  {Bennett}, \citenamefont {Kowalewski}, \citenamefont {Rouxel},\ and\
  \citenamefont {Mukamel}}]{bennett2018}%
  \BibitemOpen
  \bibfield  {author} {\bibinfo {author} {\bibfnamefont {K.}~\bibnamefont
  {Bennett}}, \bibinfo {author} {\bibfnamefont {M.}~\bibnamefont {Kowalewski}},
  \bibinfo {author} {\bibfnamefont {J.~R.}\ \bibnamefont {Rouxel}},\ and\
  \bibinfo {author} {\bibfnamefont {S.}~\bibnamefont {Mukamel}},\ }\href
  {https://doi.org/10.1073/pnas.1805335115} {\bibfield  {journal} {\bibinfo
  {journal} {Proc. Natl. Acad. Sci.}\ }\textbf {\bibinfo {volume} {115}},\
  \bibinfo {pages} {6538} (\bibinfo {year} {2018})}\BibitemShut {NoStop}%
\bibitem [{\citenamefont {Simmermacher}\ \emph {et~al.}(2019)\citenamefont
  {Simmermacher}, \citenamefont {Moreno~Carrascosa}, \citenamefont
  {E.~Henriksen}, \citenamefont {B.~M{\o}ller},\ and\ \citenamefont
  {Kirrander}}]{simmermacher2019}%
  \BibitemOpen
  \bibfield  {author} {\bibinfo {author} {\bibfnamefont {M.}~\bibnamefont
  {Simmermacher}}, \bibinfo {author} {\bibfnamefont {A.}~\bibnamefont
  {Moreno~Carrascosa}}, \bibinfo {author} {\bibfnamefont {N.}~\bibnamefont
  {E.~Henriksen}}, \bibinfo {author} {\bibfnamefont {K.}~\bibnamefont
  {B.~M{\o}ller}},\ and\ \bibinfo {author} {\bibfnamefont {A.}~\bibnamefont
  {Kirrander}},\ }\href {https://doi.org/10.1063/1.5110040} {\bibfield
  {journal} {\bibinfo  {journal} {J. Chem. Phys.}\ }\textbf {\bibinfo {volume}
  {151}},\ \bibinfo {pages} {174302} (\bibinfo {year} {2019})}\BibitemShut
  {NoStop}%
\bibitem [{\citenamefont {Giri}\ \emph {et~al.}(2022)\citenamefont {Giri},
  \citenamefont {Tremblay},\ and\ \citenamefont {Dixit}}]{giri2022}%
  \BibitemOpen
  \bibfield  {author} {\bibinfo {author} {\bibfnamefont {S.}~\bibnamefont
  {Giri}}, \bibinfo {author} {\bibfnamefont {J.~C.}\ \bibnamefont {Tremblay}},\
  and\ \bibinfo {author} {\bibfnamefont {G.}~\bibnamefont {Dixit}},\ }\href
  {https://doi.org/10.1103/PhysRevA.106.033120} {\bibfield  {journal} {\bibinfo
   {journal} {Phys. Rev. A}\ }\textbf {\bibinfo {volume} {106}},\ \bibinfo
  {pages} {033120} (\bibinfo {year} {2022})}\BibitemShut {NoStop}%
\bibitem [{\citenamefont {Tremblay}\ \emph {et~al.}(2023)\citenamefont
  {Tremblay}, \citenamefont {Blanc}, \citenamefont {Krause}, \citenamefont
  {Giri},\ and\ \citenamefont {Dixit}}]{tremblay2023a}%
  \BibitemOpen
  \bibfield  {author} {\bibinfo {author} {\bibfnamefont {J.~C.}\ \bibnamefont
  {Tremblay}}, \bibinfo {author} {\bibfnamefont {A.}~\bibnamefont {Blanc}},
  \bibinfo {author} {\bibfnamefont {P.}~\bibnamefont {Krause}}, \bibinfo
  {author} {\bibfnamefont {S.}~\bibnamefont {Giri}},\ and\ \bibinfo {author}
  {\bibfnamefont {G.}~\bibnamefont {Dixit}},\ }\href
  {https://doi.org/10.1002/cphc.202200463} {\bibfield  {journal} {\bibinfo
  {journal} {ChemPhysChem}\ }\textbf {\bibinfo {volume} {24}},\ \bibinfo
  {pages} {e202200463} (\bibinfo {year} {2023})}\BibitemShut {NoStop}%
\bibitem [{\citenamefont {Friedrich}\ \emph {et~al.}(1912)\citenamefont
  {Friedrich}, \citenamefont {Knipping},\ and\ \citenamefont {{von
  Laue}}}]{vonLaue1912}%
  \BibitemOpen
  \bibfield  {author} {\bibinfo {author} {\bibfnamefont {W.}~\bibnamefont
  {Friedrich}}, \bibinfo {author} {\bibfnamefont {P.}~\bibnamefont
  {Knipping}},\ and\ \bibinfo {author} {\bibfnamefont {M.}~\bibnamefont {{von
  Laue}}},\ }\href@noop {} {\emph {\bibinfo {title}
  {Interferenz-{{Erscheinungen}} Bei {{R{\"o}ntgenstrahlen}}}}}\ (\bibinfo
  {publisher} {Sitzungsberichte der Mathematisch-Physikalischen Classe der
  K{\"o}niglich-Bayerischen Akademie der Wissenschaften zu M{\"u}nchen},\
  \bibinfo {address} {M{\"u}nchen},\ \bibinfo {year} {1912})\BibitemShut
  {NoStop}%
\bibitem [{\citenamefont {Jones}(2014)}]{jones2014}%
  \BibitemOpen
  \bibfield  {author} {\bibinfo {author} {\bibfnamefont {N.}~\bibnamefont
  {Jones}},\ }\href {https://doi.org/10.1038/505602a} {\bibfield  {journal}
  {\bibinfo  {journal} {Nature}\ }\textbf {\bibinfo {volume} {505}},\ \bibinfo
  {pages} {602} (\bibinfo {year} {2014})}\BibitemShut {NoStop}%
\bibitem [{\citenamefont {Castro~Neto}\ \emph {et~al.}(2009)\citenamefont
  {Castro~Neto}, \citenamefont {Guinea}, \citenamefont {Peres}, \citenamefont
  {Novoselov},\ and\ \citenamefont {Geim}}]{castroneto2009}%
  \BibitemOpen
  \bibfield  {author} {\bibinfo {author} {\bibfnamefont {A.~H.}\ \bibnamefont
  {Castro~Neto}}, \bibinfo {author} {\bibfnamefont {F.}~\bibnamefont {Guinea}},
  \bibinfo {author} {\bibfnamefont {N.~M.~R.}\ \bibnamefont {Peres}}, \bibinfo
  {author} {\bibfnamefont {K.~S.}\ \bibnamefont {Novoselov}},\ and\ \bibinfo
  {author} {\bibfnamefont {A.~K.}\ \bibnamefont {Geim}},\ }\href
  {https://doi.org/10.1103/RevModPhys.81.109} {\bibfield  {journal} {\bibinfo
  {journal} {Rev. Mod. Phys.}\ }\textbf {\bibinfo {volume} {81}},\ \bibinfo
  {pages} {109} (\bibinfo {year} {2009})}\BibitemShut {NoStop}%
\bibitem [{\citenamefont {Ishikawa}(2010)}]{ishikawa2010}%
  \BibitemOpen
  \bibfield  {author} {\bibinfo {author} {\bibfnamefont {K.~L.}\ \bibnamefont
  {Ishikawa}},\ }\href {https://doi.org/10.1103/PhysRevB.82.201402} {\bibfield
  {journal} {\bibinfo  {journal} {Phys. Rev. B}\ }\textbf {\bibinfo {volume}
  {82}},\ \bibinfo {pages} {201402} (\bibinfo {year} {2010})}\BibitemShut
  {NoStop}%
\bibitem [{\citenamefont {Kelardeh}\ \emph {et~al.}(2015)\citenamefont
  {Kelardeh}, \citenamefont {Apalkov},\ and\ \citenamefont
  {Stockman}}]{kelardeh2015}%
  \BibitemOpen
  \bibfield  {author} {\bibinfo {author} {\bibfnamefont {H.~K.}\ \bibnamefont
  {Kelardeh}}, \bibinfo {author} {\bibfnamefont {V.}~\bibnamefont {Apalkov}},\
  and\ \bibinfo {author} {\bibfnamefont {M.~I.}\ \bibnamefont {Stockman}},\
  }\href {https://doi.org/10.1103/PhysRevB.91.045439} {\bibfield  {journal}
  {\bibinfo  {journal} {Phys. Rev. B}\ }\textbf {\bibinfo {volume} {91}},\
  \bibinfo {pages} {045439} (\bibinfo {year} {2015})}\BibitemShut {NoStop}%
\bibitem [{\citenamefont {Liu}\ \emph {et~al.}(2018)\citenamefont {Liu},
  \citenamefont {Manz}, \citenamefont {Ohmori}, \citenamefont {Sommer},
  \citenamefont {Takei}, \citenamefont {Tremblay},\ and\ \citenamefont
  {Zhang}}]{liu2018}%
  \BibitemOpen
  \bibfield  {author} {\bibinfo {author} {\bibfnamefont {C.}~\bibnamefont
  {Liu}}, \bibinfo {author} {\bibfnamefont {J.}~\bibnamefont {Manz}}, \bibinfo
  {author} {\bibfnamefont {K.}~\bibnamefont {Ohmori}}, \bibinfo {author}
  {\bibfnamefont {C.}~\bibnamefont {Sommer}}, \bibinfo {author} {\bibfnamefont
  {N.}~\bibnamefont {Takei}}, \bibinfo {author} {\bibfnamefont {J.~C.}\
  \bibnamefont {Tremblay}},\ and\ \bibinfo {author} {\bibfnamefont
  {Y.}~\bibnamefont {Zhang}},\ }\bibfield  {journal} {\bibinfo  {journal}
  {Phys. Rev. Lett.}\ }\textbf {\bibinfo {volume} {121}},\ \href
  {https://doi.org/10.1103/PhysRevLett.121.173201}
  {10.1103/PhysRevLett.121.173201} (\bibinfo {year} {2018})\BibitemShut
  {NoStop}%
\bibitem [{\citenamefont {Morimoto}\ \emph {et~al.}(2022)\citenamefont
  {Morimoto}, \citenamefont {Shinohara}, \citenamefont {Ishikawa},\ and\
  \citenamefont {Hommelhoff}}]{morimoto2022}%
  \BibitemOpen
  \bibfield  {author} {\bibinfo {author} {\bibfnamefont {Y.}~\bibnamefont
  {Morimoto}}, \bibinfo {author} {\bibfnamefont {Y.}~\bibnamefont {Shinohara}},
  \bibinfo {author} {\bibfnamefont {K.~L.}\ \bibnamefont {Ishikawa}},\ and\
  \bibinfo {author} {\bibfnamefont {P.}~\bibnamefont {Hommelhoff}},\ }\href
  {https://doi.org/10.1088/1367-2630/ac5c18} {\bibfield  {journal} {\bibinfo
  {journal} {New J. Phys.}\ }\textbf {\bibinfo {volume} {24}},\ \bibinfo
  {pages} {033051} (\bibinfo {year} {2022})}\BibitemShut {NoStop}%
\bibitem [{\citenamefont {Houston}(1940)}]{houston1940}%
  \BibitemOpen
  \bibfield  {author} {\bibinfo {author} {\bibfnamefont {W.~V.}\ \bibnamefont
  {Houston}},\ }\href {https://doi.org/10.1103/PhysRev.57.184} {\bibfield
  {journal} {\bibinfo  {journal} {Phys. Rev.}\ }\textbf {\bibinfo {volume}
  {57}},\ \bibinfo {pages} {184} (\bibinfo {year} {1940})}\BibitemShut
  {NoStop}%
\bibitem [{\citenamefont {Krieger}\ and\ \citenamefont
  {Iafrate}(1986)}]{krieger1986}%
  \BibitemOpen
  \bibfield  {author} {\bibinfo {author} {\bibfnamefont {J.~B.}\ \bibnamefont
  {Krieger}}\ and\ \bibinfo {author} {\bibfnamefont {G.~J.}\ \bibnamefont
  {Iafrate}},\ }\href {https://doi.org/10.1103/PhysRevB.33.5494} {\bibfield
  {journal} {\bibinfo  {journal} {Phys. Rev. B}\ }\textbf {\bibinfo {volume}
  {33}},\ \bibinfo {pages} {5494} (\bibinfo {year} {1986})}\BibitemShut
  {NoStop}%
\bibitem [{\citenamefont {Higuchi}\ \emph {et~al.}(2017)\citenamefont
  {Higuchi}, \citenamefont {Heide}, \citenamefont {Ullmann}, \citenamefont
  {Weber},\ and\ \citenamefont {Hommelhoff}}]{higuchi2017}%
  \BibitemOpen
  \bibfield  {author} {\bibinfo {author} {\bibfnamefont {T.}~\bibnamefont
  {Higuchi}}, \bibinfo {author} {\bibfnamefont {C.}~\bibnamefont {Heide}},
  \bibinfo {author} {\bibfnamefont {K.}~\bibnamefont {Ullmann}}, \bibinfo
  {author} {\bibfnamefont {H.~B.}\ \bibnamefont {Weber}},\ and\ \bibinfo
  {author} {\bibfnamefont {P.}~\bibnamefont {Hommelhoff}},\ }\href
  {https://doi.org/10.1038/nature23900} {\bibfield  {journal} {\bibinfo
  {journal} {Nature}\ }\textbf {\bibinfo {volume} {550}},\ \bibinfo {pages}
  {224} (\bibinfo {year} {2017})}\BibitemShut {NoStop}%
\bibitem [{\citenamefont {Ivakhnenko}\ \emph {et~al.}(2023)\citenamefont
  {Ivakhnenko}, \citenamefont {Shevchenko},\ and\ \citenamefont
  {Nori}}]{ivakhnenko2023}%
  \BibitemOpen
  \bibfield  {author} {\bibinfo {author} {\bibfnamefont {O.~V.}\ \bibnamefont
  {Ivakhnenko}}, \bibinfo {author} {\bibfnamefont {S.~N.}\ \bibnamefont
  {Shevchenko}},\ and\ \bibinfo {author} {\bibfnamefont {F.}~\bibnamefont
  {Nori}},\ }\href {https://doi.org/10.1016/j.physrep.2022.10.002} {\bibfield
  {journal} {\bibinfo  {journal} {Phys. Rep.}\ }\textbf {\bibinfo {volume}
  {995}},\ \bibinfo {pages} {1} (\bibinfo {year} {2023})}\BibitemShut {NoStop}%
\end{thebibliography}

%

\end{document}


\title{Supplementary Material for ``Attosecond Diffraction Imaging of Electron Dynamics in Solids''}%

\author{Mingrui Yuan}
\affiliation{Wyant College of Optical Sciences, University of Arizona, Tucson, AZ 85721}
\affiliation{Department of Physics, University of Arizona, Tucson, AZ 85721}

\author{Nikolay V. Golubev}
\email{ngolubev@arizona.edu}
\affiliation{Department of Physics, University of Arizona, Tucson, AZ 85721}

\date{\today}

\begin{abstract}
In the Letter, we present an application of the time-resolved diffraction imaging (TRDI) technique to probe the laser-driven electron dynamics in neutral graphene. We demonstrate a significant difference between the diffraction signals obtained with the semiclassical (SC) and quantum mechanical (QM) versions of the TRDI. Here, we present a step-by-step derivation of the equations used in the main text as well as the details of the performed simulations. In Sec.~\ref{sec:SM_electron_dynamics}, we discuss the general formalism for describing the electronic structure and the light-driven electron dynamics in solid state systems. Section~\ref{sec:SM_TBM_graphene} contains the analytical expressions for the band energies, related eigenstates, and the corresponding electric dipole matrix elements of graphene obtained employing the tight-binding model. In Sec.~\ref{sec:SM_SC_TRDI}, we summarize briefly the SC-TRDI approach, while Sec.~\ref{sec:SM_QM_TRDI} contains the detailed derivations of the QM-TRDI equations. In Sec.~\ref{sec:SM_diffr_graphene} we include the details of the computational procedures utilized for simulating the electron dynamics and the corresponding diffraction imaging in graphene. Section~\ref{sec:SM_diffr_abinitio} contains the comparison of graphene's electronic structure and diffraction obtained using the tight-binding model and those retrieved from first-principles electronic-structure calculations. Finally, in Sec.~\ref{sec:SM_comparison} we present comparison of our SC and QM-TRDI approaches with previously reported results on diffraction imaging of electron dynamics in graphene. 
\end{abstract}

\pacs{Valid PACS appear here}
\maketitle

\newpage

\section{Laser-driven electron dynamics in solids}
\label{sec:SM_electron_dynamics}
In this section, we present a general tight-binding model for simulating the light-driven electron dynamics in solid state systems. We use atomic units for all derivations below for simplicity. Let us consider a periodic crystal lattice formed by, in general non-equivalent, atoms each having a set of corresponding atomic orbitals. In our calculations we neglect the motion of atomic nuclei focusing instead on the dynamics of the electrons which, in turn, can be described using the so-called Bloch states~\cite{bloch1929}
\begin{equation}
\label{eq:Bloch_states}
	\Phi_m(\mathbf{k},\mathbf{r}) = \frac{1}{\sqrt{N}} \sum_{i=1}^{N} e^{i\mathbf{k} \cdot \mathbf{R}_{m,i}} 
		\varphi_m(\mathbf{r}-\mathbf{R}_{m,i}),
\end{equation}
where index $m$ denotes a particular orbital located at position $\mathbf{R}_{m,i}$ in $i$-th unit cell, and $\mathbf{k}$ and $\mathbf{r}$ are the vectors in the momentum and position spaces, respectively. The total number of the generated Bloch states, which are linear superpositions of atomic orbitals of $N$ unit cells, is equal to the number $M$ of atomic orbitals present in each unit cell. We assume the atomic orbitals are normalized: $\int |\varphi_m(\mathbf{r})|^2 d\mathbf{r} = 1$, so that a single electron occupies each orbital.

The field-free electronic eigenstates $\Psi_n(\mathbf{k},\mathbf{r})$ of the crystal can be represented as a linear superposition of Bloch states
\begin{equation}
	\Psi_n(\mathbf{k},\mathbf{r})=\sum_{m=1}^{M} \psi_{n,m}(\mathbf{k}) \Phi_m(\mathbf{k},\mathbf{r}),
\end{equation}
where $\psi_{n,m}(\mathbf{k})$ are the expansion coefficients. To simplify the follow-up discussion, we utilize the following standard notation:
\begin{equation}
	\Psi_n(\mathbf{k},\mathbf{r}) = \langle \mathbf{r} | \Psi_n(\mathbf{k}) \rangle
		=\sum_{m=1}^{M} \psi_{n,m}(\mathbf{k}) \langle \mathbf{r} | \Phi_m(\mathbf{k}) \rangle,
\end{equation}
i.e. we assume the brackets $|\quad\rangle$ represent the integration over $\mathbf{r}$ space. Accordingly, the expansion coefficients $\psi_{n,m}(\mathbf{k})$ are obtained by minimizing the energies
\begin{equation}
	E_n(\mathbf{k})= \frac{\langle \Psi_n(\mathbf{k})| \hat{H}_0 |\Psi_n(\mathbf{k}) \rangle}
		{\langle \Psi_n(\mathbf{k}) | \Psi_n(\mathbf{k}) \rangle}
\end{equation}
of the field-free states $|\Psi_n(\mathbf{k}) \rangle$ for a system described by $\hat{H}_0$ Hamiltonian. Applying the variational theorem, the evaluation of the expansion coefficients boils down to solving the following generalized eigenvalue problem~\cite{slater1954}
\begin{equation}
\label{eq:GEP}
	H_0(\mathbf{k}) \psi_n (\mathbf{k}) = E_n(\mathbf{k}) S(\mathbf{k}) \psi_n (\mathbf{k}),
\end{equation}
where $\psi^T_n=(\psi_{n,1},\psi_{n,2},...,\psi_{n,M})$ is a vector composed from the corresponding expansion coefficients, and matrices $H_0(\mathbf{k})$ and $S(\mathbf{k})$ are representation of the Hamiltonian and the overlap integrals, respectively, in the basis of Bloch states:
\begin{equation}
	H_{0,\mu,\nu}(\mathbf{k})=\langle \Phi_{\mu}(\mathbf{k}) | \hat{H}_0 | \Phi_{\nu}(\mathbf{k}) \rangle
	\qquad \text{and} \qquad
	S_{\mu,\nu}(\mathbf{k})=\langle \Phi_{\mu}(\mathbf{k}) | \Phi_{\nu}(\mathbf{k}) \rangle.
\end{equation}
In the nearest-neighbor approximation (see, e.g., Ref.~\cite{harrison1989}), the tight-binding Hamiltonian for electrons is an operator acting in the basis of atomic orbitals:
\begin{equation}
\label{eq:TBH}
	\hat{H}_0 = -t \sum_{\langle i,j \rangle,\sigma}(a^{\dagger}_{i,\sigma} a^{}_{j,\sigma} + \text{H.c.}),
\end{equation}
where $a^{\dagger}_{i,\sigma}$ and $a^{}_{j,\sigma}$ are creation and annihilation operators, respectively, indices $i$ and $j$ run over atomic orbitals, $\sigma$ is the spin polarization, $t$ is the hopping energy, and $\text{H.c.}$ denotes the Hermitian conjugate. Therefore, solving Eq.~(\ref{eq:GEP}), one obtains the band energies $E_n(\mathbf{k})$ and the corresponding wavefunctions $\psi_n (\mathbf{k})$ of the tight-binding Hamiltonian, Eq.~(\ref{eq:TBH}). The obtained quantities, in turn, can be used to describe the structure and dynamics of electrons in a periodic crystal in the presence of external electromagnetic fields.

The interaction of a solid state system with an external field can be described by the time-dependent Hamiltonian
\begin{equation}
\label{eq:td_H}
	\hat{H}(t) = \hat{H}_0 - i \mathbf{A}(t) \cdot \nabla + \frac{1}{2}A^2(t),
\end{equation}
where $\mathbf{A}(t)=-\int_{-\infty}^{t}\mathbf{E}(t')dt'$ is the vector potential of the applied electric field $\mathbf{E}(t)$. Solving the time-dependent Schr\"odinger equation with the Hamiltonian~(\ref{eq:td_H}) and employing the dipole approximation, one can express the time-dependent electronic wavefunction of the system as
\begin{equation}
\label{eq:electronic_wave_packet}
	|\Psi (\mathbf{k},t) \rangle=\sum_{n=1}^{M} c_n(\mathbf{k},t) | \Psi_n(\mathbf{k}_t) \rangle,
\end{equation}
where $c_n(\mathbf{k},t)$ are expansion coefficients that reflect the evolution of the populations of the time-dependent electronic eigenstates $|\Psi_n(\mathbf{k}_t)\rangle$, such that
\begin{equation}
    \langle \mathbf{r} | \Psi_n(\mathbf{k}_t) \rangle = \exp \left( -i \mathbf{A}(t) \cdot \mathbf{r} \right)
    \Psi_n(\mathbf{k}_t,\mathbf{r}),
\end{equation}
and $\mathbf{k}_t=\mathbf{k}_0+\mathbf{A}(t)$ is the crystal momentum frame evolving according to the Bloch acceleration theorem~\cite{houston1940}.

By introducing the density matrix $\rho_{n,m}(\mathbf{k},t)=c_n(\mathbf{k},t) c^{*}_m(\mathbf{k},t)$, the time evolution of the expansion coefficients can be described by semiconductor Bloch equation~\cite{krieger1986}
\begin{equation}
\label{eq:SBEs}
\begin{aligned}
  i \frac{\partial }{\partial t} \rho_{n,m}(\mathbf{k},t) & =
  	\left[E_m(\mathbf{k}_t) - E_n(\mathbf{k}_t)\right] \rho_{n,m}(\mathbf{k},t) \\ 
  & + \mathbf{E}(t) \cdot \left\{ \mathbf{D}(\mathbf{k}_t),\mathbf{\rho}(\mathbf{k},t)\right\}_{n,m}
  	-i\frac{1-\delta_{n,m}}{T_d} \rho_{n,m}(\mathbf{k},t), \\
\end{aligned}
\end{equation}
where $T_d$ is the interband dephasing time, the commutator symbol ``$\{\}$'' is defined as $\{A,B\}=AB-BA$, and $\mathbf{D}\left( \mathbf{k} \right) = \left( \mathbf{D}_x(\mathbf{k}),\mathbf{D}_y(\mathbf{k}),\mathbf{D}_z(\mathbf{k}) \right)^T$ is a vector of matrices of the transition dipole moments connecting the bands with each other. The required dipole moments can be obtained from the following well-known relation~\cite{gu2013}:
\begin{equation}
\label{eq:dipoles}
	\left( D_{x;n,m}(\mathbf{k}),D_{y;n,m}(\mathbf{k}),D_{z;n,m}(\mathbf{k}) \right)^T
		=\frac{i}{E_m(\mathbf{k})-E_n(\mathbf{k})}
		\langle \Psi_n(\mathbf{k})| S^{-1}(\mathbf{k}) \nabla_\mathbf{k} H(\mathbf{k})| \Psi_m(\mathbf{k})\rangle.
\end{equation}

To visualize the laser-induced electron dynamics in real space, we compute the electron density
\begin{equation}
	Q(\mathbf{r},t)=\frac{1}{\mathcal{N}}
		\int_{\text{unit cell}}
			\langle \Psi(\mathbf{k},t) | \hat{\rho}(\mathbf{r}) | \Psi(\mathbf{k},t) \rangle
		d\mathbf{k},
\end{equation}
where $\hat{\rho}(\mathbf{r})=\delta(\mathbf{r}-\mathbf{r}')$ is the local density operator and the integration is performed over the unit cell in the reciprocal space. The normalization factor $\mathcal{N}$ is chosen such that the integral of the real-space electron density over the unit cell in the real space
\begin{equation}
	\int_{\text{unit cell}} Q(\mathbf{r},t) d\mathbf{r} = M
\end{equation}
returns the number of active electrons $M$. Accordingly, $\mathcal{N}=\mathcal{V}/M$, where $\mathcal{V}$ is the volume of the unit cell in the reciprocal space.

Utilizing the expansion of the time-dependent wavefunction $| \Psi(\mathbf{k},t) \rangle$ in the basis of electronic eigenstates, Eq.~(\ref{eq:electronic_wave_packet}), we can express the electron density as
\begin{equation}
\label{eq:Q_total}
	Q(\mathbf{r},t)=\frac{1}{\mathcal{N}} \sum_{n,m} \int_{\text{unit cell}} \rho_{n,m}(\mathbf{k},t) Q_{m,n} (\mathbf{k}_t,\mathbf{r}) d \mathbf{k},
\end{equation}
where
\begin{equation}
\label{eq:Qmn}
	Q_{m,n} (\mathbf{k},\mathbf{r}) = \Psi_m^* (\mathbf{k},\mathbf{r}) \Psi_n (\mathbf{k},\mathbf{r})
\end{equation}
denote the $\mathbf{k}$-resolved electron densities of eigenstates when $m=n$, or the transition densities between the states when $m \neq n$. The total electron density $Q(\mathbf{r},t)$ can be split in two components: $Q(\mathbf{r},t)=Q_{\text{intra}}(\mathbf{r},t)+Q_{\text{inter}}(\mathbf{r},t)$, where the first term $Q_{\text{intra}}(\mathbf{r},t)$ represents the diagonal portion of Eq.~(\ref{eq:Q_total}) ($n=m$) and thus is referred to as the intraband electron density, while the second term $Q_{\text{inter}}(\mathbf{r},t)$ is composed from the off-diagonal elements responsible for the coherent interband electron motion.

\section{Tight-binding model of graphene}
\label{sec:SM_TBM_graphene}
\begin{figure}[t]
	\includegraphics[width=\textwidth]{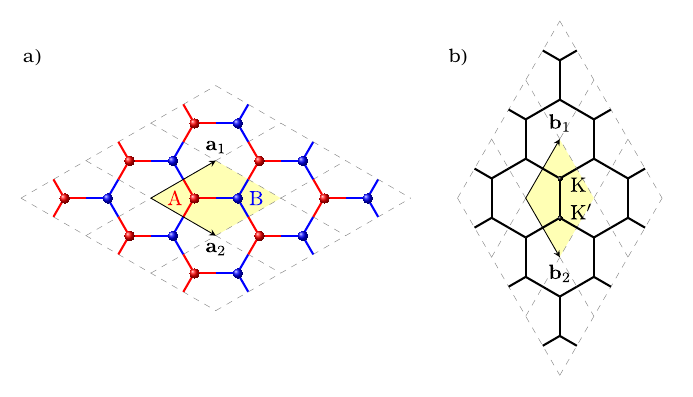}
\caption{The real (a) and reciprocal (b) space structure of graphene.}
\label{fig:graphene_lattice}
\end{figure}

Graphene is made out of carbon atoms arranged in hexagonal structure (see, e.g., Ref.~\cite{castroneto2009}), as shown in Fig.~\ref{fig:graphene_lattice}(a). The graphene unit cell, determined by two lattice vectors $\mathbf{a}_1 = \frac{a}{2}(\sqrt{3},1)$ and $\mathbf{a}_2 = \frac{a}{2}(\sqrt{3},-1)$, contains two carbon atoms ``A'' and ``B'' located at positions $a(1/\sqrt{3},0)$ and $a(2/\sqrt{3}, 0)$, respectively. Accordingly, the distance between the nearest neighbor atoms of graphene is $a/\sqrt{3}$, where $a$ = 2.46~$\AA$ is the lattice constant.

The corresponding reciprocal space lattice, shown in Fig.~\ref{fig:graphene_lattice}(b), is formed by vectors $\mathbf{b}_1=\frac{2\pi}{a}(1/\sqrt{3},1)$ and $\mathbf{b}_2=\frac{2\pi}{a}(1/\sqrt{3},-1)$. Of particular importance for the physics of graphene are the two points K and K$'$, named Dirac points, with coordinates $\frac{2\pi}{a}(1/\sqrt{3},1/3)$ and $\frac{2\pi}{a}(1/\sqrt{3},-1/3)$, respectively. These points together with their replicas in neighboring unit cells around the origin form a hexagonal first Brillouin zone of the reciprocal lattice of graphene.

Each carbon atom of graphene posses six electrons while four of them participate in the formation of valence bonds. Three of these (the so-called $\sigma$-electrons) form tight bonds with the neighboring atoms in the plane and do not participate in the dynamics initiated by the laser pulse used in our simulations. The fourth electron (the so-called $\pi$-electron), instead, can be excited and then accelerated by the laser pulse which can manifest in the change of the electron density of the system in time. Accordingly, we construct the Bloch states of graphene from $2p_z$ atomic orbitals of carbon atoms~\cite{stroucken2011}:
\begin{equation}
\label{eq:2pz_orb}
	\varphi_{2p_z}(\mathbf{r})=\frac{\sqrt{Z^5}}{4\sqrt{2\pi}}
		\mathbf{r} \cos(\theta) \exp \left(-\frac{Z}{2} \mathbf{r} \right),
\end{equation}
where $Z$ is the effective nuclear charge, $\theta$ is the polar angle of $\mathbf{r}$ measured from the z-axis, and the orbitals are assumed to be located at the atomic nucleus. Utilizing the orbital defined by Eq.~(\ref{eq:2pz_orb}) and substituting the corresponding atomic positions to Eq.~(\ref{eq:Bloch_states}), we obtain the following representation of the tight-binding Hamiltonian and the overlap matrix in the basis of Bloch states:
\begin{equation}
\label{eq:H0_S_graphene}
	H_0(\mathbf{k})=
	\begin{pmatrix}
		\varepsilon_{2p_z} & \gamma f(\mathbf{k}) \\
		\gamma f(\mathbf{k})^{*} & \varepsilon_{2p_z}
	\end{pmatrix}
	\qquad \text{and} \qquad
	S(\mathbf{k})=
	\begin{pmatrix}
		1 & s f(\mathbf{k}) \\
		s f(\mathbf{k})^{*} & 1
	\end{pmatrix},
\end{equation}
where
\begin{equation}
	\varepsilon_{2p_z}=\int \varphi^*_{2p_z}(\mathbf{r}) \hat{H}_0 \varphi_{2p_z}(\mathbf{r}) d\mathbf{r}
\end{equation}
is the energy of the $2p_z$ orbital,
\begin{equation}    
    \gamma=\int \varphi^*_{2p_z}(\mathbf{r}-\mathbf{R}_A) \hat{H}_0 \varphi_{2p_z}(\mathbf{r}-\mathbf{R}_B) d\mathbf{r}
\end{equation}
is the hopping integral between two neighboring orbitals of carbon atoms, and
\begin{equation}    
    s=\int \varphi^*_{2p_z}(\mathbf{r}-\mathbf{R}_A) \varphi_{2p_z}(\mathbf{r}-\mathbf{R}_B) d\mathbf{r}
\end{equation}
is the overlap integral between those orbitals. The function $f(\mathbf{k})$ describes the spatial part of the nearest-neighbor hopping for honeycomb structure of graphene:
\begin{equation}  
    f(\mathbf{k})=\exp \left( \frac{i k_x a}{\sqrt{3}} \right)+2 \exp \left( -\frac{i k_x a}{2\sqrt{3}} \right)\cos \left( \frac{k_y a}{2} \right).
\end{equation}

The secular Eq.~(\ref{eq:GEP}) for the Hamiltonian and overlap matrix defined by Eqs.~(\ref{eq:H0_S_graphene}) can be solved analytically, giving thus access to the band energies
\begin{equation}
\label{eq:graphene_E}
	E_{n}(\mathbf{k})=
	\frac{\varepsilon_{2p_z} \pm \gamma |f(\mathbf{k})|}
		{1 \pm s |f(\mathbf{k})|},
\end{equation}
and the corresponding reciprocal-space eigenvectors of graphene
\begin{equation}
	\psi_{n}(\mathbf{k}) = \frac{1}{\sqrt{2(1 \pm s |f(\mathbf{k})|)}}
	\begin{pmatrix}
		1 \\
		\pm e^{-i \phi(\mathbf{k})}
	\end{pmatrix},
\end{equation}
where the index $n \in \{v,c\}$ denotes the valence ($v$) and the conduction ($c$) bands, and $\phi(\mathbf{k})=\text{Arg}[f(\mathbf{k})]$ is the phase of $f(\mathbf{k})$. Furthermore, Eq.~(\ref{eq:dipoles}) gives access to the transition dipole moments between the valence and conduction bands of graphene which can be expressed as
\begin{equation}
\label{eq:graphene_Dx}
	D_{cv,x}(\mathbf{k}) = \frac{1}{2\sqrt{3}} \frac{a}{|f(\mathbf{k})|^2} 
		\sqrt{\frac{1-s|f(\mathbf{k})|}{1+s|f(\mathbf{k})|}} \left\{
			\cos(ak_y) - \cos\left( \frac{\sqrt{3} a k_x}{2}\right) \cos\left( \frac{a k_y}{2}\right)
		\right\}
\end{equation}
for $x$ component, and
\begin{equation}
\label{eq:graphene_Dy}
	D_{cv,y}(\mathbf{k}) = \frac{1}{2} \frac{a}{|f(\mathbf{k})|^2} 
		\sqrt{\frac{1-s|f(\mathbf{k})|}{1+s|f(\mathbf{k})|}}
			\sin\left( \frac{\sqrt{3} a k_x}{2}\right) \sin\left( \frac{a k_y}{2}\right),
\end{equation}
for the corresponding $y$ component.

\section{Semiclassical theory of diffraction imaging}
\label{sec:SM_SC_TRDI}
A traditional approach to describe the scattering of X-rays or electrons from matter employs a semiclassical approximation, wherein matter is treated as a quantum object while the incident and scattered beams are described in the frame of classical electrodynamics. In the first Born approximation, the interaction between the beam and the system is assumed to be weak and no excited states are involved in the scattering process. Accordingly, the differential scattering probability of waves that come out in the solid angle $d\Omega$ can be expressed as the Fourier transform of the electron density $Q(\mathbf{r})$ of the system~\cite{ashcroft1976,coppens1992,coppens1997}
\begin{equation}
\label{eq:classical_DSP}
    \frac{d P}{d \Omega}=\frac{d P_e}{d \Omega} \left|
    	\int Q(\mathbf{r}) e^{i \mathbf{S} \cdot \mathbf{r}} d \mathbf{r}
    \right|^2 = \frac{d P_e}{d \Omega} \left| \hat{\mathcal{F}}_{\mathbf{S}}[Q(\mathbf{r})] \right|^2,
\end{equation}
where $d P_e /d \Omega$ is the differential scattering probability from a free electron, $\mathbf{S}=\mathbf{k}_{\text{in}} - \mathbf{k}_{\text{s}}$ is the momentum transfer with $\mathbf{k}_{\text{in}}$ and $\mathbf{k}_{\text{s}}$ being the momentum of the incident and scattered beams, respectively, and $\hat{\mathcal{F}}_{\mathbf{S}}$ denotes the Fourier transform operator defined for the given scattering vector $\mathbf{S}$. In the following, we will utilize the scattering intensity $I(\mathbf{S})=\left| \hat{\mathcal{F}}_{\mathbf{S}}[Q(\mathbf{r})] \right|^2$, in units of the scattering from a free electron, instead of the differential scattering probability for convenience.

In case of scattering on the periodic crystal, the electron density $Q(\mathbf{r})$ is a three-dimensional periodic function that can be expressed in a form of a convolution as~\cite{coppens1992,coppens1997}
\begin{equation}
\label{eq:Q_convolution}
    Q_{\text{crystal}}(\mathbf{r}) = \sum_n \sum_m \sum_p \int_{-\infty}^{+\infty} 
    	Q_{\text{unit cell}}(\mathbf{r}') \delta(\mathbf{r}-\mathbf{r}'-n\mathbf{a}_1-m\mathbf{a}_2-p\mathbf{a}_3) d\mathbf{r}',
\end{equation}
where $Q_{\text{unit cell}}(\mathbf{r})$ is the electron density within the unit cell, $n$, $m$, and $p$ are integers, $\mathbf{a}_1$, $\mathbf{a}_2$, and $\mathbf{a}_3$ are real-space lattice vectors of the crystal, and $\delta$ denotes the Dirac delta function. According to the Fourier convolution theorem, the Fourier transform of a convolution of two functions can be expressed as the product of the Fourier transforms of the individual functions:
\begin{equation}
\label{eq:diffr_convolution}
	\hat{\mathcal{F}}_{\mathbf{S}}[Q_{\text{crystal}}(\mathbf{r})] = 
		\hat{\mathcal{F}}_{\mathbf{S}}[Q_{\text{unit cell}}(\mathbf{r})] 
		\sum_h \sum_k \sum_l \delta(\mathbf{S}-h\mathbf{b}_1-k\mathbf{b}_2-l\mathbf{b}_3),
\end{equation}
where $h$, $k$, and $l$ are new set of integers, also called Miller indices~\cite{ashcroft1976}, and $\mathbf{b}_1$, $\mathbf{b}_2$, and $\mathbf{b}_3$ are the reciprocal-space vectors. As one can see from Eq.~(\ref{eq:diffr_convolution}), the diffraction from a crystal will show up in a form of a characteristic diffraction pattern with the scattered beams being observed at specific positions $\mathbf{S}$ defined by combinations of the reciprocal space vectors:
\begin{equation}
\label{eq:Svector}
	\mathbf{S} = h\mathbf{b}_1+k\mathbf{b}_2+l\mathbf{b}_3,
\end{equation}
and having the intensities
\begin{equation}
\label{eq:classical_diffr_amplitude}
	I(\mathbf{S})= \left|
		\int_{\text{unit cell}} Q(\mathbf{r}) e^{i \mathbf{S} \cdot \mathbf{r}} d \mathbf{r}
	\right|^2.
\end{equation}

When the system evolves in time, one could expect that the electron density will also be a time-dependent quantity. Accordingly, a straightforward approach to the time-resolved diffraction imaging will be to plug $Q(\mathbf{r},t)$ instead of the stationary density $Q(\mathbf{r})$ in Eq.~(\ref{eq:classical_diffr_amplitude}). Taking the electron density from Eq.~(\ref{eq:Q_total}) and substituting it to Eq.~(\ref{eq:classical_diffr_amplitude}), we obtain the following expression for the time-dependent scattering intensities:
\begin{equation}
\label{eq:I_SC}
	I_{\text{SC}}(\mathbf{S},t)=
		\frac{1}{\mathcal{N}} 
		\left| 
			\sum_{n,m} \int_{\text{unit cell}} \rho_{n,m}(\mathbf{k},t)
			\mathcal{\hat{F}}_{\mathbf{S}}[Q_{m,n} (\mathbf{k}_t,\mathbf{r})]
			d\mathbf{k}
		\right|^2,
\end{equation}
where the subscript SC refers to the semiclassical formalism used for deriving this expression.

\section{Quantum theory of diffraction imaging}
\label{sec:SM_QM_TRDI}
In their pioneering work~\cite{dixit2012}, Dixit and co-workers demonstrated that physically correct description of the time-resolved diffraction imaging requires explicit consideration of mechanisms responsible for the scattering from a time-dependent target. In this section, we utilize Eq.~(S16) from Supporting Information of Ref.~\cite{dixit2012} (see also Eq.~(5) from Ref.~\cite{dixit2014}) as the starting point in developing a fully quantum theory of the time-resolved diffraction imaging of the electron dynamics in solids. The time-dependent differential scattering probability can be expressed as~\cite{dixit2012,dixit2014}
\begin{equation}
\label{eq:DSP_QM}
\begin{aligned}
	\frac{d P(t)}{d \Omega} = & \frac{d P_e}{d \Omega} 
		\int_0^{\infty} d \omega_{\mathbf{k}_{\text{s}}} W_{\Delta E} \left(\omega_{\mathbf{k}_{\text{s}}}\right) 
			\frac{\omega_{\mathbf{k}_{\text{s}}}}{\omega_{\mathbf{k}_{\text{in}}}}
		\int_{-\infty}^{\infty} \frac{d \tau}{2 \pi} C(\tau) 
			e^{-i(\omega_{\mathbf{k}_{\text{s}}}-\omega_{\mathbf{k}_{\text{in}}}) \tau} \\
        & \times \int d \mathbf{r} \int d \mathbf{r}^{\prime}
        	\left\langle \Psi \left|
        		\hat{\rho}\left(\mathbf{r}^{\prime}, t+\frac{\tau}{2}\right) 
        		\hat{\rho}\left(\mathbf{r},          t-\frac{\tau}{2}\right)
        	\right| \Psi \right\rangle 
        e^{i \mathbf{S} \cdot (\mathbf{r}-\mathbf{r}^{\prime})},
\end{aligned}
\end{equation}
where $W_{\Delta E} \left(\omega_{\mathbf{k}_{\text{s}}}\right)$ is a spectral window function, $C(\tau)=\exp (-2 \ln 2 \tau^2 / \sigma^2)$ is a function of the probe pulse duration $\sigma$, $|\Psi\rangle$ is the time-independent wavefunction, and $\hat{\rho}(\mathbf{r},t)$ is the time-dependent electron density operator in the Heisenberg representation. As one can see, Eq.~(\ref{eq:DSP_QM}) represents the double Fourier transform (integrals over $\mathbf{r}$ and $\mathbf{r}'$ with $ e^{i \mathbf{S} \cdot (\mathbf{r}-\mathbf{r}^{\prime})}$ factor) of the so-called density-density correlation function (term in $\langle \quad \rangle$ brackets) where, in addition, the finite duration of the probe pulse is taken into account (integral over $\tau$ and $C(\tau)$ term) and also the fact that the scattering can be inelastic (integral over $\omega_{\mathbf{k}_{\text{s}}}$ and the terms containing the energies of the incident $\omega_{\mathbf{k}_{\text{in}}}$ and the scattered $\omega_{\mathbf{k}_{\text{s}}}$ photons). Furthermore, the energy resolution of the scattering detector is also included by the presence of the $W_{\Delta E} \left(\omega_{\mathbf{k}_{\text{s}}}\right)$ function, which models the sensitivity of the detector in the $\Delta E$ energy range to the incoming photons of $\omega_{\mathbf{k}_{\text{s}}}$ energy.

To simplify Eq.~(\ref{eq:DSP_QM}), we assume that the duration of the probe pulse is much shorter than the time scale of the electron dynamics under study. In the limit of the infinitely short probe pulse, we can write
\begin{equation}
	\lim_{\sigma \to 0} C(\tau) = 
	\lim_{\sigma \to 0} \exp{\left(-2 \ln 2 \frac{\tau^2}{\sigma^2}\right)} = 
	\delta(\tau),
\end{equation}
thus obtaining a delta function. We assume, in addition, that the energy of the incident and the scattered photons are comparable to each other: $\omega_{\mathbf{k}_{\text{in}}} \approx \omega_{\mathbf{k}_{\text{s}}}$, and that the detector cannot resolve them: $W_{\Delta E} \left(\omega_{\mathbf{k}_{\text{s}}}\right) \approx 1$. Importantly, we do not exclude the inelastic scattering processes from consideration since we do not modify the density-density correlation function responsible for them. Instead, we rely on a hypothesis that the intensities of the scattered beams themselves contain enough signatures of the dynamics of the system neglecting an additional information one could extract measuring the change in the photon energies. Applying the simplifications discussed above to Eq.~(\ref{eq:DSP_QM}), the intensities of the scattered beams can be expressed as
\begin{equation}
\label{eq:I_QM_full}
    I_{\text{QM}}(\mathbf{S},t)=
    \int d \mathbf{r} \int d \mathbf{r}^{\prime}
        	\left\langle \Psi \left|
        		\hat{\rho}\left(\mathbf{r}^{\prime}, t\right) 
        		\hat{\rho}\left(\mathbf{r},          t\right)
        	\right| \Psi \right\rangle 
        e^{i \mathbf{S} \cdot (\mathbf{r}-\mathbf{r}^{\prime})},
\end{equation}
where the subscript QM is introduced to distinguish the new equation from the SC results discussed in the previous section.

To proceed forward, let us switch from the Heisenberg to the Schr\"odinger picture by expressing the time-dependent density operators as
\begin{equation}
\label{eq:rho_SH}
	\hat{\rho}(\mathbf{r},t) = \hat{U}^\dagger(t) \hat{\rho}(\mathbf{r}) \hat{U}(t),
\end{equation}
where $\hat{U}(t)=\hat{T} e^{-i \int_{-\infty}^t d\tau \hat{H}(\tau)}$ is the time-evolution operator induced by, in general time-dependent, Hamiltonian $\hat{H}(t)$ and $\hat{\rho}(\mathbf{r})$ is the time-independent density operator in the Schr\"odinger representation. Substituting Eq.~(\ref{eq:rho_SH}) to Eq.~(\ref{eq:I_QM_full}), we obtain
\begin{equation}
    I_{\text{QM}}(\mathbf{S},t)=
    \int d \mathbf{r} \int d \mathbf{r}^{\prime}
        	\left\langle \Psi (t) \left|
        		\hat{\rho}(\mathbf{r}^{\prime}) 
        		\hat{\rho}(\mathbf{r})
        	\right| \Psi(t) \right\rangle 
        e^{i \mathbf{S} \cdot (\mathbf{r}-\mathbf{r}^{\prime})},
\end{equation}
where we utilized the unitarity of the evolution operator: $\hat{U}^\dagger(t) \hat{U}(t) \equiv 1$, and made the wavefunction time-dependent: $\hat{U}(t) |\Psi \rangle = |\Psi(t) \rangle$.

Next, we switch from a general wavefunction $|\Psi(t)\rangle$ to the representation Eq.~(\ref{eq:electronic_wave_packet}), thus obtaining
\begin{equation}
\label{eq:I_QM_sums_ints}
\begin{split}
    I_{\text{QM}}(\mathbf{S},t)= & \sum_{n,m}
    \int d \mathbf{r} \int d \mathbf{r}^{\prime}
    \int d \mathbf{k} \quad
        c^{*}_m(\mathbf{k},t) c_n(\mathbf{k},t) \\
        & \times \langle \Psi_m (\mathbf{k}_t)|
        	\hat{\rho}(\mathbf{r}^{\prime})
        	\sum_f  |\Psi_f (\mathbf{k}_t) \rangle \langle \Psi_f (\mathbf{k}_t)|
        	\hat{\rho}(\mathbf{r})
        |\Psi_n (\mathbf{k}_t) \rangle
        e^{i \mathbf{S} \cdot (\mathbf{r}-\mathbf{r}^{\prime})},
\end{split}
\end{equation}
where we inserted, in addition, the resolution of identity $\sum_f  |\Psi_f (\mathbf{k}_t) \rangle \langle \Psi_f (\mathbf{k}_t)| \equiv 1$ between the density operators. Rearranging the integrals and summations in Eq.~(\ref{eq:I_QM_sums_ints}), we finally obtain
\begin{equation}
\label{eq:I_QM_final}
\begin{split}
	I_{\text{QM}}(\mathbf{S},t) = & \sum_{n,m,f}
	\int d \mathbf{k} \rho_{n,m}(\mathbf{k},t) 
		\int Q_{m,f}(\mathbf{k}_t,\mathbf{r}') e^{-i \mathbf{S} \cdot \mathbf{r}^{\prime}} d \mathbf{r}^{\prime} 
		\int Q_{f,n}(\mathbf{k}_t,\mathbf{r})  e^{ i \mathbf{S} \cdot \mathbf{r}} d \mathbf{r} \\
	& = \frac{1}{\mathcal{N}} \sum_{n,m,f}
	\int_{\text{unit cell}} \rho_{n,m}(\mathbf{k},t)
		\mathcal{\hat{F}}^{*}_{\mathbf{S}}[Q_{f,m} (\mathbf{k}_t,\mathbf{r})]
		\mathcal{\hat{F}}_{\mathbf{S}}[Q_{f,n} (\mathbf{k}_t, \mathbf{r})]
	d\mathbf{k},
\end{split}
\end{equation}
where we utilized the fact that $c^{*}_m(\mathbf{k},t) c_n(\mathbf{k},t)=\rho_{n,m}(\mathbf{k},t)$ are the matrix elements of the reciprocal space density matrix, $Q_{m,f}=Q^*_{f,m}$ are defined by Eq.~(\ref{eq:Qmn}), integrals over $\mathbf{r}$ and $\mathbf{r}'$ are separable, and the Fourier transform has the following property:
\begin{equation}
	\int Q_{m,f}(\mathbf{k}_t,\mathbf{r}') e^{-i \mathbf{S} \cdot \mathbf{r}'} d \mathbf{r}' 
	= \left(
		\int Q^{*}_{m,f}(\mathbf{k}_t,\mathbf{r}') e^{i \mathbf{S} \cdot \mathbf{r}'} d \mathbf{r}'
	  \right)^*
	= \mathcal{\hat{F}}^{*}_{\mathbf{S}}[Q_{f,m} (\mathbf{k}_t,\mathbf{r})].
\end{equation}
Furthermore, we utilized the periodicity of $\mathbf{k}$- and $\mathbf{r}$-spaces thus replacing the corresponding integrals and Fourier transforms by those performed within the unit cell. The derived Eq.~(\ref{eq:I_QM_final}) is composed from physically intuitive terms and can be easily compared to a similar expression, Eq.~(\ref{eq:I_SC}), obtained employing the SC formalism.

\section{Imaging of electron dynamics in graphene}
\label{sec:SM_diffr_graphene}
We simulated the temporal evolution of the electron density in graphene driven by the interaction with the laser field of the following waveform:
\begin{equation}
	\mathbf{E}(t)=\mathbf{e} E_0 \sin^4(\pi t/\tau)\cos(\omega t),
		\qquad \text{for $|t| \le \tau$ and 0 otherwise}, 
\end{equation}
where the peak field amplitude $E_0$ is chosen to be 2.5~V/nm, pulse duration $\tau$ is 21~fs, the photon energy $\omega$ is 1.55~eV which corresponds to the wavelength of 800~nm, and $\mathbf{e}$ denotes the unit vector in the direction of the field polarization. The evolution of the density matrix $\mathbf{\rho }(\mathbf{k},t)$ in reciprocal space is obtained by numerically solving Eq.~(\ref{eq:SBEs}). We sampled the unit cell in the first Brillouin zone of graphene with a uniform $257\times 257$ grid along the reciprocal lattice vectors. The required energies of the bands $E_n(\mathbf{k})$ and the transition dipole moments $\mathbf{D}(\mathbf{k})$ are obtained from Eqs.~(\ref{eq:graphene_E}) and (\ref{eq:graphene_Dx}--\ref{eq:graphene_Dy}). The integration in the time domain is performed by the Runge--Kutta--Fehlberg method~\cite{fehlberg1969} with adaptive time step control. We used $\varepsilon_{2p_z}=0$ for the orbital energy, and $s=0.127$ for the overlap integral, which corresponds to the effective nuclear charge $Z=4.02$. The lattice parameter of graphene $a$ is set to be 2.46~$\AA$ and the dephasing time $T_d$ is 10~fs. The real-space electron densities and the corresponding diffraction intensities are computed using the uniform $16\times16$ grid sampling the unit cell along the lattice vectors. The $z$ axis is sampled by 16 uniformly distributed grid points along the perpendicular direction passing through the graphene sheet and covering the distance from 0 to 3~$\AA$.

\begin{figure}[h!]
	\includegraphics[width=0.9\textwidth]{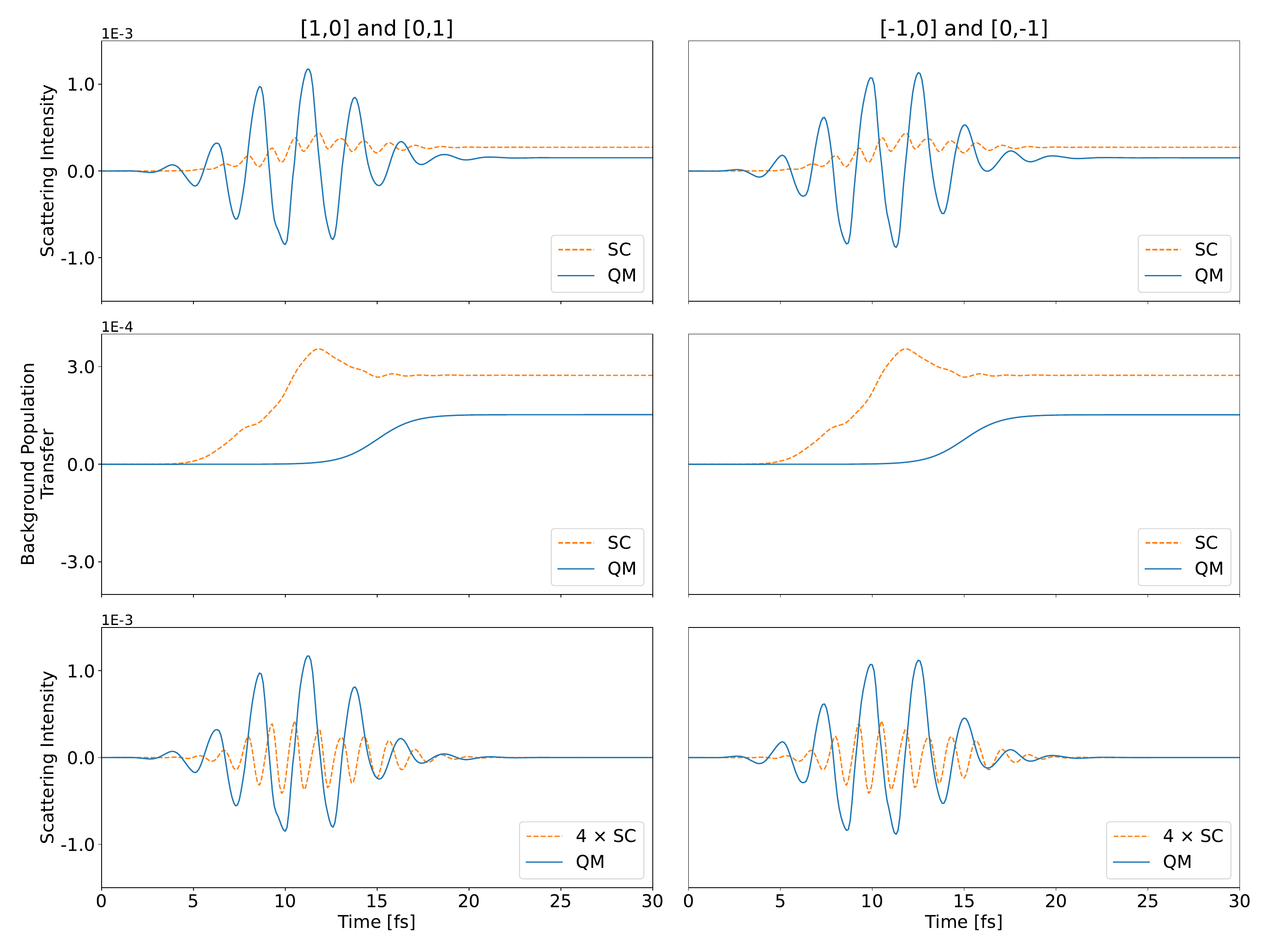}
\caption{Comparison of the time-resolved Bragg diffraction intensities in graphene (left column: $[1,0]$ and $[0,1]$ spots, right column: $[-1,0]$ and $[0,-1]$ spots) computed with the semiclassical (SC, orange dashed lines) and quantum mechanical (QM, blue solid lines) approaches. Top panel: Diffraction intensities with the subtracted static background. Middle panel: Extracted population transfer background. Bottom panel: Oscillating portion of the diffraction signal obtained by removing the static and population transfer backgrounds from the calculated intensities. Note that the SC intensities in the bottom panel are scaled by a factor of four for better visibility.}
\label{fig:raw_diffr_10}
\end{figure}
\begin{figure}[t]
	\includegraphics[width=0.9\textwidth]{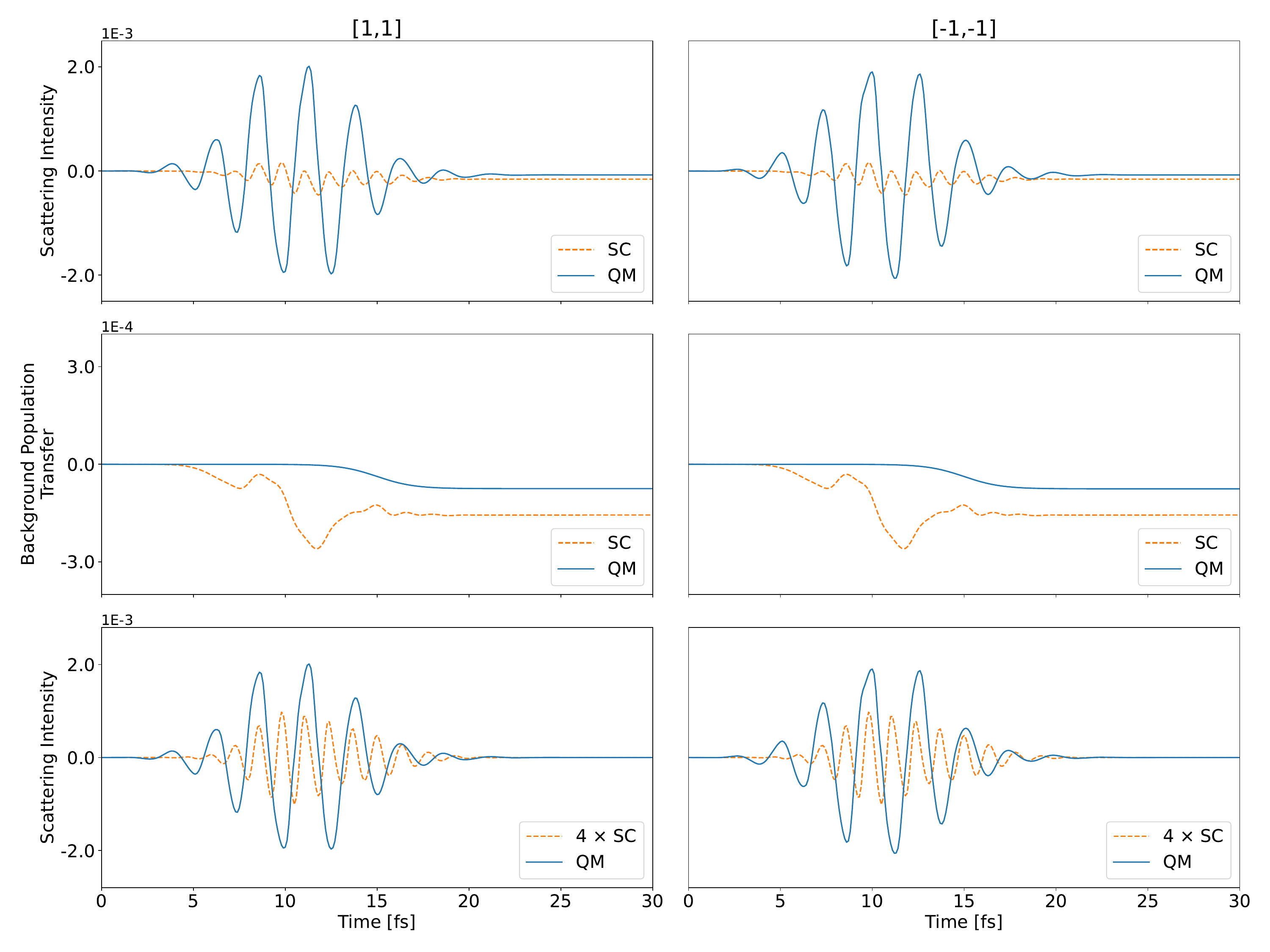}
\caption{Same analysis as that presented in Fig.~\ref{fig:raw_diffr_10} performed for $[1,1]$ (left column) and $[-1,-1]$ (right column) Bragg diffraction spots of graphene.}
\label{fig:raw_diffr_11}
\end{figure}

We computed the SC and QM diffraction signals defined by Eqs.~(\ref{eq:I_SC}) and~(\ref{eq:I_QM_final}), respectively. Importantly, the SC and QM diffraction intensities are not equal to each other even at the initial moment of time when the laser pulse does not act on the system. This is due to the fact that the SC approach, Eq.~(\ref{eq:I_SC}), includes the elastic scattering only, while the QM expression, Eq.~(\ref{eq:I_QM_final}), takes the inelastic processes into account. Furthermore, we found that the magnitude of the oscillations of the diffraction signal in the SC approach is approximately four times weaker in comparison to the QM results. In addition, the diffraction signal contains a contribution responsible for the population transfer which show up differently for the SC and QM simulations. To simplify the comparison of the SC and QM results, we processed the corresponding intensities by removing the static portion of the signal and also the contribution responsible for the populations transfer. In case of the SC approach, we used a standard Fourier filtering technique to extract the oscillating part of the signal. For the QM results, the contribution of the population transfer is much weaker which makes it hard to extract it using the Fourier filter. Instead, we assume the population transfer occurs smoothly between the initial and final values of the signal and extract this background from the calculated intensities. Figures~\ref{fig:raw_diffr_10} and~\ref{fig:raw_diffr_11} depict the calculated intensities (top panel) together with the extracted population transfer background (middle panel) and the resulting normalized signals (bottom panel) for all six first-order Bragg diffraction peaks of graphene. We would like to emphasize that the normalization procedure does not influence the main conclusion of the paper regarding the frequency of the observed oscillations. The only purpose of the employed data processing is to make the comparison between the SC and QM results more clear.

\begin{figure}[t]
	\includegraphics[width=\textwidth]{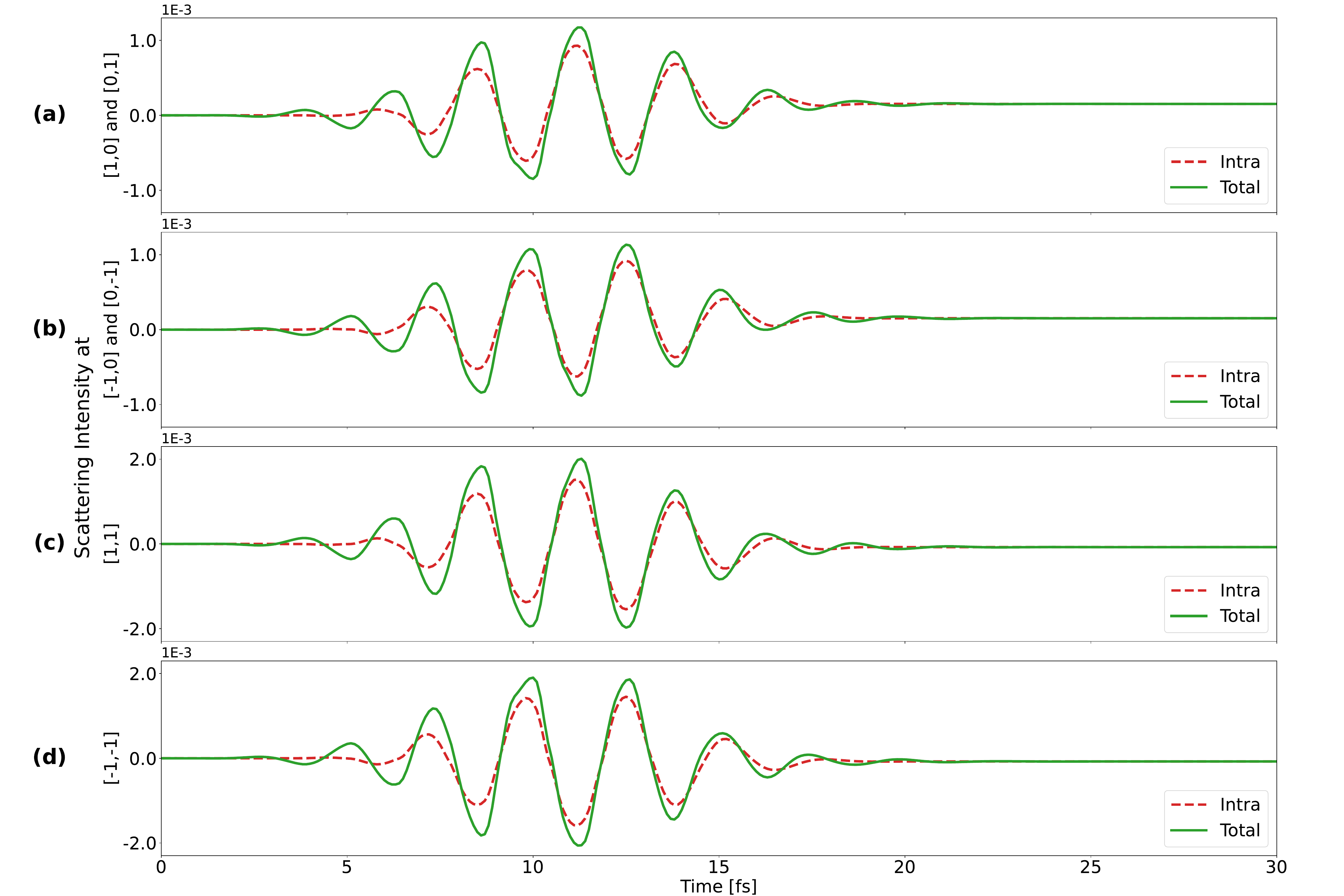}
\caption{Intraband (red dashed lines) and total (green solid lines) QM scattering intensities computed for all six first-order Bragg diffraction spots of graphene.}
\label{fig:diffr_intra_total}
\end{figure}

As we mentioned in the main text, the linearity of Eq.~(\ref{eq:I_QM_final}) makes it possible to separate the contributions to the diffraction signal arising from the intra- and interband electron dynamics. Figure~\ref{fig:diffr_intra_total} illustrates the total QM diffraction signal (green solid lines) as well as the intraband contribution (red dashed lines) computed for all six first-order Bragg diffraction peaks of graphene. As one can see, in all cases the total signal is composed almost entirely from the intraband contribution with a small portion of the overall intensity coming from the interband dynamics.


%

\section{Diffraction imaging employing first-principles electronic-structure calculations}
\label{sec:SM_diffr_abinitio}
One of the differences between the conventional semiclassical and our fully quantum approaches to the description of the diffraction process is the inclusion of the inelastic scattering pathways to the cross section. The QM-TRDI expression, Eq.~(\ref{eq:I_QM_final}) [Eq.~3 in the main text], contains a summation over three indices which correspond to the pairs of initial states ($n$ and $m$) forming the electronic wave packet and the final state $f$ reached by the scattering process. Accordingly, in the quantum treatment of the diffraction we account for the possibility that a photon or an electron of the probe beam can exchange energy with the electrons of a target leading to the fact that the target will end up in a different state after the scattering event. It is, therefore, important to analyze the convergence of Eq.~(\ref{eq:I_QM_final}) with respect to the number of final states participating in the scattering process.

\begin{figure}[t]
	\includegraphics[width=0.9\textwidth]{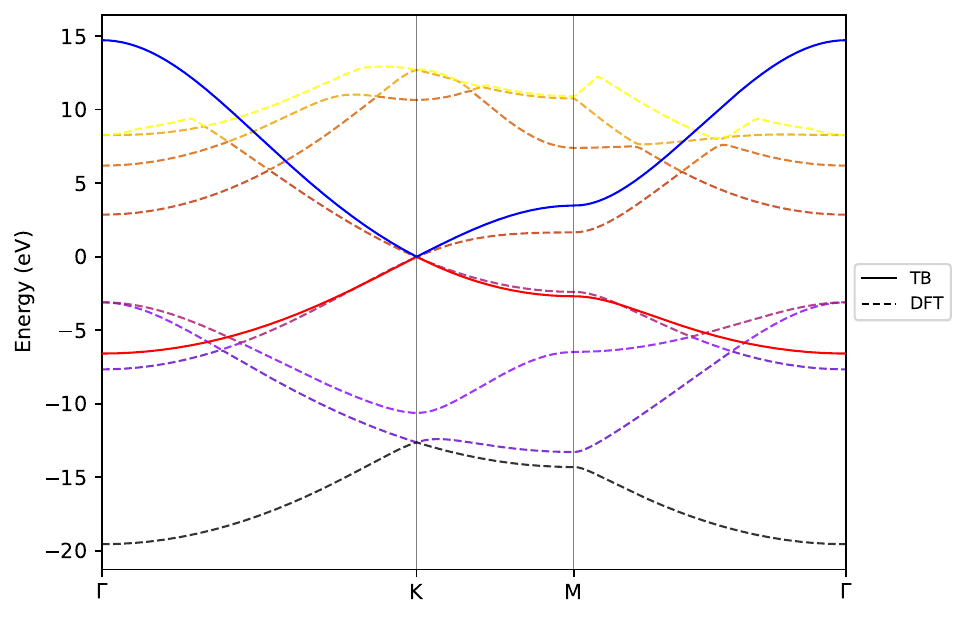}
\caption{Comparison of the band structure of graphene in reciprocal space obtained using the tight-binding (TB) and DFT approaches.}
\label{fig:TB_DFT_bands}
\end{figure}

To go beyond a simple two-band tight-binding model of graphene presented in Sec.~\ref{sec:SM_TBM_graphene}, we performed first-principles electronic-structure calculations using density functional theory (DFT) approach in the framework of the plane-wave pseudopotential method. The calculations were performed using powerful Quantum ESPRESSO package~\cite{QE2009,QE2017}. We utilized a standard ultrasoft scalar relativistic pseudopotential of carbon atoms (Kresse--Joubert projector-augmented-wave (PAW) potential parameterized using Perdew--Burke--Ernzerhof (PBE) exchange-correlation functional, C.pbe-n-kjpaw\_psl.1.0.0.UPF, taken from the Quantum ESPRESSO pseudopotential data base~\cite{PPQElibrary}) assuming two 1s electrons of each atom remain frozen.

Our simulations consisted of several stages. At the first step, we performed the ground state self-consistent field calculations of Kohn--Sham orbitals on a uniform 129$\times$129 reciprocal space grid covering the unit cell (\textbf{pw.x} module of the Quantum ESPRESSO). Then the obtained ground state wavefunctions have been used to calculate the reciprocal space wavefunctions of excited states and the corresponding energies of the bands (\textbf{pw.x} module with ``nbnd'' keyword). Finally, we recovered the real-space wavefunctions for every band at every $\mathbf{k}$ point (\textbf{wfck2r.x} module) and thus obtained the required quantities to represent the transition densities $Q_{m,n} (\mathbf{k},\mathbf{r})$ using Eq.~(\ref{eq:Qmn}). We included eight bands of graphene in our simulations. The calculated band structure is shown in Fig.~\ref{fig:TB_DFT_bands} where we also plot the band energies obtained using the tight-binding model, Eqs.~(\ref{eq:graphene_E}), for comparison. As one can see, the energies of $\pi$ and $\pi^*$ bands of graphene are in a good agreement between two models, especially at the vicinity of the Dirac point $K$ where the studied dynamics takes place. The inelastic scattering to the other bands not included in the two-level tight-binding model of graphene could potentially lead to extra features in the diffraction signal that we analyze next.

\begin{figure}[t]
	\includegraphics[width=0.9\textwidth]{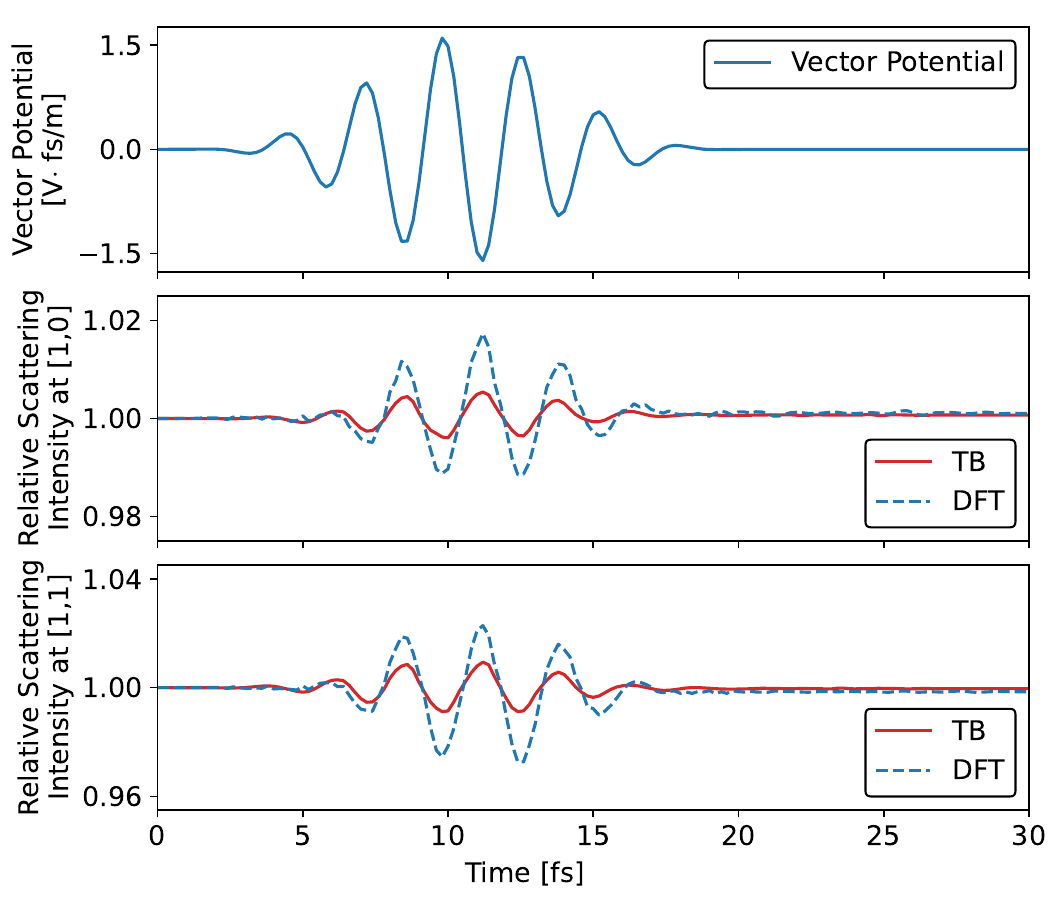}
\caption{Comparison of the QM-TRDI intensities in graphene computed using the two-band tight-binding (TB) model and eight-band first-principles density functional theory (DFT) approach. Top panel: Vector potential of the applied electric field (blue solid line). Middle and bottom panels: relative diffraction intensities at $[1,0]$ and $[1,1]$ Bragg spots, respectively, computed using the TB (red solid lines) and DFT (blue dashed lines) approaches.}
\label{fig:TB_DFT_comparison}
\end{figure}

For computing the diffraction intensities, the procedure described in the previous paragraph has been executed for each propagation time step on a shifted reciprocal space grid $\mathbf{k}_t$ giving thus access to the time-dependent transition densities $Q_{m,n} (\mathbf{k}_t,\mathbf{r})$ present in Eq.~(\ref{eq:I_QM_final}) for the QM-TRDI. Figure~\ref{fig:TB_DFT_comparison} depicts the comparison of the QM-TRDI intensities obtained using the two-band tight-binding model and eight-band first-principles DFT approach. As one can see, adding more bands into the simulations somewhat increases the scattering intensities. However, the dynamical features of the diffraction signals such as the oscillation period and the overall shape of the curves are almost identical to each other in both cases. Accordingly, we conclude that the main physics of the time-resolved diffraction is contained in the matrix elements $\rho_{n,m}(\mathbf{k},t)$ of the reciprocal space density matrix while the real-space transition densities $Q_{m,n} (\mathbf{k}_t,\mathbf{r})$ are responsible for the intensity of the corresponding transitions. It is, of course, desirable to include as many bands as possible into the simulations to achieve the convergent results. However, the time-dependence and characteristic features of the diffraction intensities are fully captured by a simple two-level tight-binding model of graphene. We note in passing that the convergence of the QM-TRDI approach has been analyzed already for the case of molecules where fast convergence with respect to the number of final states included into simulations has been demonstrated~\cite{hermann2020}.

Importantly, the conclusions made in this section are valid for the specific pump laser pulse parameters used in our simulations. More intense or energetic laser pulses may induce transitions between the bands not considered in the two-band tight-binding approach. In this case, the use of the general methodology presented in this section is fully justified. Furthermore, the DFT approach and the utilized Quantum ESPRESSO package can be easily used for describing properties of an arbitrary periodic solid thus making the presented QM-TRDI formalism immediately applicable to various systems of theoretical and/or experimental interest.

\section{Comparison of SC- and QM-TRDI approaches with previous calculations of diffraction in graphene}
\label{sec:SM_comparison}
We mentioned in the main text that Yakovlev and co-workers demonstrated theoretically an application of the SC-TRDI to measure the ultrafast laser-driven electron dynamics in graphene~\cite{yakovlev2015}. In a difference to our study, the authors of Ref.~\cite{yakovlev2015} utilized a phenomenological lattice potential to describe the interaction of electrons in graphene and express the time-dependent Schr\"odinger equation in three spatial dimensions. The advantage of this approach over the tight-binding model is that the lattice potential can support inner electrons that could be important for diffraction. On the other hand, the use of the reciprocal space density matrix, Eq.~(\ref{eq:SBEs}), gives access to the physically intuitive visual representation of the electron dynamics between electronic bands of solids as shown in the panel (b) of Fig.~1 in the main text. We note in passing that many observable properties of the electron dynamics in solids such as the high-harmonic generation (HHG) spectra and the macroscopic electron currents can be extracted from the reciprocal space density matrix without the necessity to perform a way more expensive real-space calculations. Furthermore, the reciprocal space density matrix approach is, in principle, not bound to the tight-binding model and can be easily extended to support an arbitrary representation of wavefunctions of the corresponding bands. This extensibility of our technique is demonstrated in Sec.~\ref{sec:SM_diffr_abinitio} of this document where we utilized the DFT approach to obtain the required quantities, including those corresponding to the inner electrons of carbon atoms.

To quantify the difference between our approach and the model used by Yakovlev \textit{et al.}, we performed calculations of the electron dynamics in graphene using our tight-binding model with the field taken from Ref.~\cite{yakovlev2015}. The utilized laser field is defined by the following vector potential:
\begin{equation}
\label{eq:field_yak}
	\mathbf{A}(t)=-\mathbf{e}_y \frac{E_0}{\omega} \cos^4 \left( \frac{\pi t}{2\tau} \right) \sin(\omega t), 
		\qquad \text{for $|t| \le \tau$ and 0 otherwise}, 
\end{equation}
where the photon energy $\omega$ corresponds to a central wavelength $\lambda=2\pi c/\omega =$ 2.1 $\mu$m, the full width at half maximum (FWHM) duration is 10 fs, the peak field $E_0$ is 0.5 V/$\AA$, and the polarization vector $\mathbf{e}_y$ is chosen to be perpendicular to the C-C bonds of graphene ($y$ direction in our simulations). Other parameters of the simulations are taken to be the same as presented in Sec.~\ref{sec:SM_diffr_graphene} of this document.

After calculating the evolution of the reciprocal space density matrix and the corresponding real space electron density, we can, in principle, evaluate the intensity of diffraction spots using Eqs.~(\ref{eq:I_SC}) and~(\ref{eq:I_QM_final}). However, to make an explicit comparison between our results and those presented in Ref.~\cite{yakovlev2015} we processed the obtained data as following:
\begin{itemize}
	\item Instead of performing the Fourier transform of the transition densities $Q_{m,n} (\mathbf{k},\mathbf{r})$ in $(x,y)$-plane and then integrating over $z$ direction, we computed the required $\mathcal{\hat{F}}_{\mathbf{S}}[Q_{m,n} (\mathbf{k},\mathbf{r})]$ terms as following:
\begin{equation}
	\mathcal{\hat{F}}_{\mathbf{S}}[Q_{m,n} (\mathbf{k},\mathbf{r})]
		=\frac{4\pi}{S^2+k_z^2} \int_z e^{i k_z z} \iint_{x,y \in \text{unit cell}} Q_{m,n} (\mathbf{k},\mathbf{r}) 
			e^{i \mathbf{S} \cdot \mathbf{r}}  dxdydz
		=\tilde{\varphi}^{m,n}_{\mathbf{S}}(\mathbf{k},t),
\end{equation}
thus accounting for the changes of $p_z$ component of the incident beam upon scattering (see SM of Ref.~\cite{yakovlev2015}) and utilizing the fact that $\mathbf{S}$ vector has only $x$ and $y$ components for graphene. We note that the obtained quantities $\tilde{\varphi}^{m,n}_{\mathbf{S}}(\mathbf{k},t)$ together with the reciprocal space density matrix $\bm{\rho}(\mathbf{k},t)$ can be used to recover $\tilde{\varphi}_{\mathbf{S}}(k_z,t)$:
\begin{equation}
	\tilde{\varphi}_{\mathbf{S}}(k_z,t) = \sum_{n,m} \int_{k_x,k_y \in \text{unit cell}} \rho_{n,m}(\mathbf{k},t) \tilde{\varphi}^{m,n}_{\mathbf{S}}(\mathbf{k}_t,t) dk_x dk_y + \tilde{\varphi}^{\text{(core)}}_{\mathbf{S}}(k_z),
\end{equation}
given by Eq.~(7) in Ref.~\cite{yakovlev2015}, and $\tilde{\varphi}^{\text{(core)}}_{\mathbf{S}}(k_z)$ is the ionic potential screened by core electrons. Since for the given parameters of the driver field $\tilde{\varphi}^{\text{(core)}}_{\mathbf{S}}(k_z)$ represents an additive constant contribution to the scattering signal and thus does not bring any information about the electron dynamics in the system, we neglect it in our simulations.

	\item We convolved the obtained scattering probabilities with a Gaussian function:
\begin{equation}
	p_{\mathbf{S}} (\tau) = \frac{4}{k_0 \sqrt{k_0^2 - S^2}} \int_{-\infty}^{\infty} dt 
		\frac{\exp\left(-\frac{(t-\tau)^2}{T^2} \right)}{\sqrt{\pi}T}
		\left| \tilde{\varphi}_{\mathbf{S}}\left( \frac{S^2}{2k_0},t \right) \right|^2,
\end{equation}
thus accounting for the finite duration of the probe pulse and also for the uncertainty of atomic positions in the lattice due to disorder and thermal motion~\cite{yakovlev2015}. Following the discussion presented in Ref.~\cite{yakovlev2015}, we utilized an average momentum of projectile electrons $k_0 = 30$ keV and a FWHM of $2T\sqrt{\ln 2} = 1$ fs.
\end{itemize}

\begin{figure}[t]
	\includegraphics[width=0.9\textwidth]{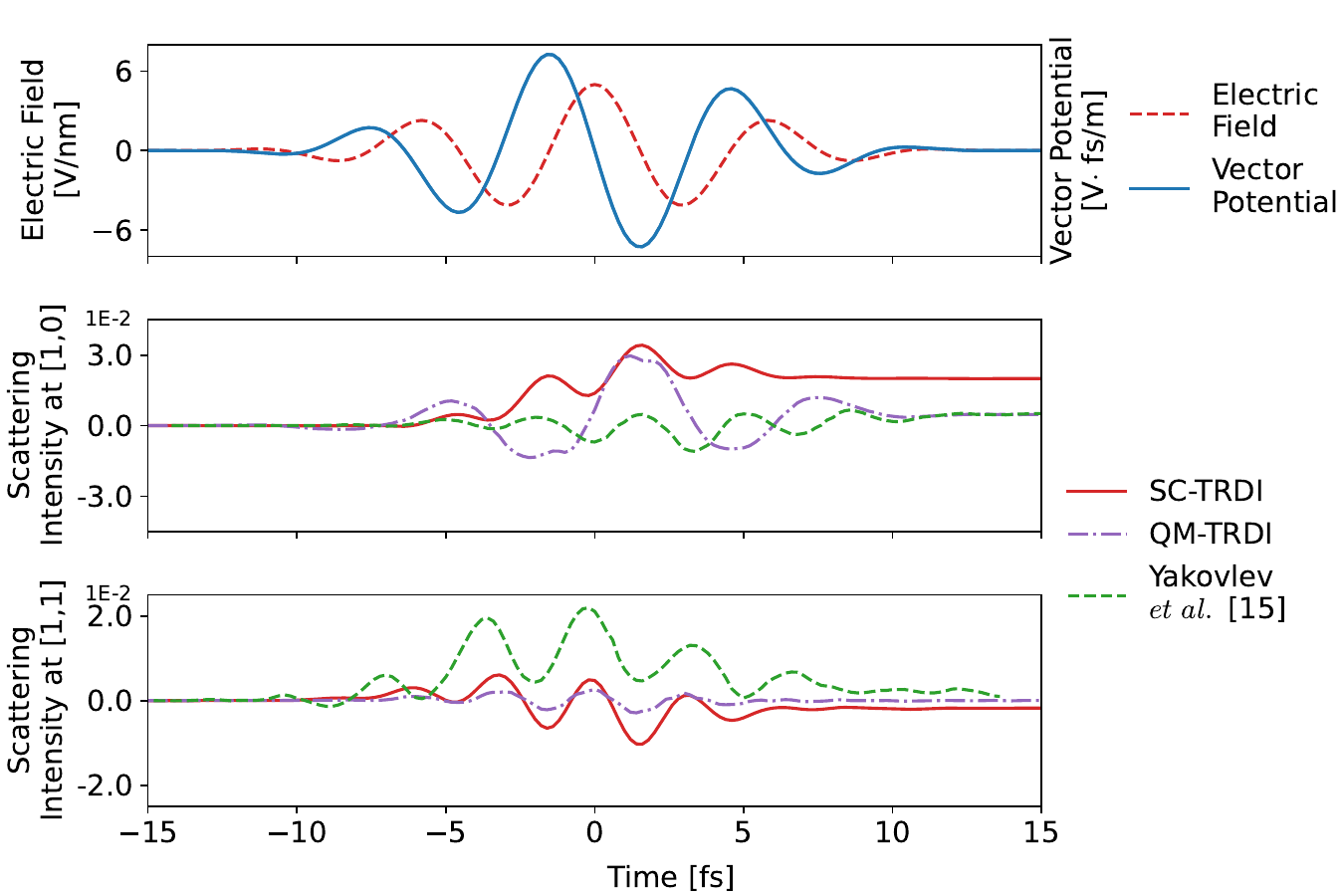}
\caption{Comparison of the time-resolved Bragg diffraction intensities in graphene computed using our SC- and QM-TRDI approaches with the semiclassical results presented in Ref.~\cite{yakovlev2015}. Top panel: Electric field (red dashed line) and the corresponding vector potential (blue solid line) of the applied laser pulse. Middle and bottom panels: SC (red solid lines) and QM (purple dash-dotted lines) diffraction intensities measured at $[1,0]$ and $[1,1]$ Bragg spots, respectively, plotted together with the corresponding signals extracted from Fig.~3 of Ref.~\cite{yakovlev2015} (green dashed lines).}
\label{fig:yak_comparison}
\end{figure}

The results of our simulations together with the data extracted from Fig.~3 of Ref.~\cite{yakovlev2015} are shown in Fig.~\ref{fig:yak_comparison}. As one can see, the phase and amplitude of oscillations of the SC-TRDI signal obtained with our model (red solid lines) agree well with the calculations from Ref.~\cite{yakovlev2015} (green dashed lines). In both cases the diffraction intensities oscillate with double the frequency of the applied driver field. The difference in the behavior of the background of the diffraction signals between our calculations and the results of Ref.~\cite{yakovlev2015} is due to the difference in the real-space structure of $\pi$ and $\pi^*$ bands of graphene. We note that in a contrast to the approach used in Ref.~\cite{yakovlev2015}, in our model the properties of the corresponding bands can be adjusted in the real and reciprocal spaces independently from each other. Accordingly, using accurate real-space densities obtained from the first-principles calculations (see Sec.~\ref{sec:SM_diffr_abinitio} of this document) one can achieve better accuracy in predictions of the diffraction signals.

Interestingly, utilizing the field given by Eq.~(\ref{eq:field_yak}), the period of oscillations of the QM-TRDI signal (purple dash-dotted lines in Fig.~\ref{fig:yak_comparison}) is found to be different for $[1,0]$ and $[1,1]$ Bragg spots of graphene. The scattering to $[1,0]$ spot demonstrates the oscillations with the same frequency as the driver field. However, the QM-TRDI measured at $[1,1]$ spot oscillates with double the frequency of the field. The reason for this distinct behavior of the two diffraction spots is due to the polarization of the applied driver field. The scattering vectors $\mathbf{S}$ of graphene corresponding to $[1,0]$ and $[1,1]$ Bragg spots are (see Eq.~(\ref{eq:Svector}) of this document):
\begin{equation}
	\mathbf{S}_{[1,0]} = 1 \mathbf{b_1} + 0 \mathbf{b_2} = \frac{2\pi}{a} \left( \frac{1}{\sqrt{3}},1 \right),
\end{equation}
and
\begin{equation}
	\mathbf{S}_{[1,1]} = 1 \mathbf{b_1} + 1 \mathbf{b_2} = \frac{2\pi}{a} \left( \frac{2}{\sqrt{3}},0 \right).
\end{equation}
We remind the reader that the main difference between the SC- and QM-TRDI approaches is the fact that the latter one captures the momentum of the moving electrons of the target. Accordingly, the electron dynamics driven by the laser pulse polarized along $x$ axis (the case considered in our paper) will be measurable at both $[1,0]$ and $[1,1]$ diffraction spots since the corresponding scattering vectors are sensitive to the $k_x$ component of the electron momentum. In a difference, if the laser pulse is polarized along $y$ axis (the case considered by Yakovlev \textit{et al.}~\cite{yakovlev2015}), the $[1,1]$ spot will not capture the motion of electrons in $y$ direction since the $\mathbf{S}$ vector of this spot does not contain the corresponding component of the momentum. In this case, the QM-TRDI signal measured at $[1,1]$ spot will not be much different from its semiclassical analogue since both approaches essentially probe the instantaneous snapshots of the time-dependent electron density which change over time due the influence of the applied laser field.

\newpage

%